\documentstyle[11pt,aaspp4]{article}

\tighten
\eqsecnum

\slugcomment{submitted to ApJ, May 9, 1996}

\begin{document}
%
\def\kms {km~s$^{-1} \,$}
\def\lsim{ \lower .75ex \hbox{$\sim$} \llap{\raise .27ex \hbox{$<$}} }
\def\gsim{ \lower .75ex \hbox{$\sim$} \llap{\raise .27ex \hbox{$>$}} }
\def\items{\hangindent=0.5truecm \hangafter=1 \noindent}
\newcommand{\vecq}{{\bf q}}
\newcommand{\vecr}{{\bf r}}
\newcommand{\vecu}{{\bf u}}
\newcommand{\vecv}{{\bf v}}
\newcommand{\vecx}{{\bf x}}
\newcommand{\Msol}{\mbox{M$_{\sun}$}}
\title{The Effects of a Photoionizing UV Background \\
on the Formation of Disk Galaxies}

\author{Julio F. Navarro\altaffilmark{1,2}}
\affil{Steward Observatory, University of Arizona, Tucson, AZ, 85721, USA.}
\and

\author{Matthias Steinmetz\altaffilmark{3}}
\affil{Max--Planck Institut f\"ur Astrophysik, Postfach 1523,
85740 Garching, Germany.\\Department of Astronomy, University of
California, Berkeley, CA 94720, USA.}

\altaffiltext{1}{Bart J.~Bok Fellow} 
\altaffiltext{2}{E-mail: jnavarro@as.arizona.edu} 
\altaffiltext{3}{E-mail: msteinmetz@astro.berkeley.edu} 

\begin{abstract}
We use high resolution N-body/gasdynamical simulations to investigate
the effects of a photoionizing UV background on the assembly of disk
galaxies in hierarchically clustering universes. We focus on the mass
and rotational properties of gas that can cool to form centrifugally
supported disks in dark matter halos of different mass. Photoheating
can significantly reduce the amount of gas that can cool in galactic
halos. Depending on the strength of the UV background field, the
amount of cooled gas can be reduced by up to $50\%$ in systems with
circular speeds in the range $80$-$200$ \kms. The magnitude of the
effect, however, is not enough to solve the ``overcooling'' problem
that plagues hierarchical models of galaxy formation if the UV
background is chosen to be consistent with estimates based on recent
observations of QSO absorption systems. Photoionization has little
effect on the collapse of gas at high redshift and affects
preferentially gas that is accreted at late times. Since disks form
inside-out, accreting higher angular momentum gas at later times,
disks formed in the presence of a UV background have spins that are
even smaller than those formed in simulations that do not include the
effects of photoionization. This exacerbates the angular momentum
problem that afflicts hierarchical models of disk formation.  We
conclude that photoionization cannot provide the heating mechanism
required to reconcile hierarchically clustering models with
observations. Energy feedback and enrichment processes from the
formation and evolution of stars must therefore be indispensable
ingredients for any successful model of the formation of disk
galaxies.
\end{abstract}


\section{Introduction} \label{intro}

Hierarchical clustering is at present the most successful paradigm of
structure formation. In this scenario, structure grows as systems of
progressively larger masses merge and collapse to form newly
virialized systems. Many aspects of this hierarchy, such as the mass
function of dark matter halos, their merger rates, and the formation
times of systems of different mass can be computed analytically using
variations of the Press-Schechter theory (Press \& Schechter 1974,
Bond  et al. 1991, Bower 1991, Lacey \& Cole 1993, Kauffmann \& White
1993). These analytical predictions have been successfully tested
against the results of N-body simulations (Lacey \& Cole 1994), and
provide a framework within which we can interpret the evolution of the
mass hierarchy in the universe once the power spectrum of initial
density fluctuations and the cosmological parameters are specified.

The formation and evolution of galaxies within this hierarchy is less
well understood. In the standard lore, galaxies form as baryons cool
and collapse in the potential wells of an evolving population of dark
matter halos (White \& Rees 1978). One key ingredient thus seems to be
the ability of gas to cool efficiently and collapse to the center of
halos, where it can form centrifugally supported disks and stellar
systems concentrated enough to survive as separate entities as their
surrounding halos merge in groups and galaxy clusters. Cooling times on
galaxy scales seem, indeed, to be at present shorter than a Hubble
time, lending support to this picture (Silk 1977, Rees
\& Ostriker 1977). Furthermore, dissipative collapse within dark halos
seems to be the only way to turn the meager angular momentum induced
by tidal torques into the fast spins of spiral galaxies (Fall and
Efstathiou 1980).

A few major shortcomings of this scenario have also been
identified. Since cooling times scale inversely with density, the
dissipative collapse of gas should have been more efficient at high
redshift because the dark matter halos present at that time (and the
universe as a whole) were denser. Cooling is expected to be so
efficient at early times that, in order to prevent the gas from
cooling and turning into stars as soon as the first level of the
hierarchy collapses, it is necessary to postulate the existence of a
heating mechanism that can keep the gas hot and diffuse until systems
with the typical size and mass of galaxies are assembled. Such heating
mechanism, if especially effective in low mass halos, may also explain
why the number of low luminosity galaxies is much lower than expected
from the steep low-mass end slope of the halo mass function. This
``overcooling'' problem and its relation to the luminosity function
was originally noticed by White \& Rees (1978) and has been explored
in detail by Cole (1991) and White \& Frenk (1991), among others.

A related problem concerns the angular momentum of gaseous disks
assembled in this scenario. In the absence of heating, most of the
mass of a galactic disk would be accreted through mergers of
protogalaxies whose own gas component had previously collapsed to form
centrifugally supported disks. Numerical simulations show that,
because of the central concentration of the gas, a large fraction of
its angular momentum is transferred to the surrounding halos during
mergers. As a result, the spin of gaseous disks formed by hierarchical
mergers is much lower than that of observed spirals (Navarro, Frenk \&
White 1995a).

These problems all suggest that an efficient heating mechanism is
needed in order to reconcile hierarchical galaxy formation scenarios
with observations. Heating may either prevent gas from collapsing in,
or remove it from the center of, low mass halos at high redshift. The
gas may thus retain its angular momentum during mergers and cool later
to form disks with spins comparable to observed spirals. The same
process also could potentially bring the galaxy luminosity function
and the halo mass function into agreement (see Kauffmann, White \&
Guiderdoni 1994, Cole et al. 1994).

One process that has been repeatedly advocated in the literature
envisions energy feedback from supernovae and evolving stars as the
main heating mechanism. This is certainly possible energetically. The
total energy released by supernovae is enough to unbind most of the
baryons from galaxy halos, especially from the shallow potential wells
of low-mass systems (see, eg., Dekel \& Silk 1986). If this energy
could be channeled efficiently into heating the surrounding
interstellar medium, perhaps driving large-scale winds such as those
seen in galaxies that are actively forming stars, they could provide
the required heating mechanism. Modeling this process is a formidable
challenge because it involves the ill-understood physics of star
formation and of the interaction of evolving stars with the
interstellar medium.

A simpler alternative has been proposed by Efstathiou (1992), who
suggested that the photoionizing UV background implied by the
Gunn-Peterson effect in QSO spectra may be responsible for preventing
gas from cooling in low mass halos at high redshift. This suggestion
can in principle be tested directly using N-body/gasdynamical
simulations. In fact, a few studies have already been carried out,
albeit with mixed results. For example, Cen \& Ostriker (1992) argued
that even a UV background much weaker than suggested by the
Gunn-Peterson effect could prevent the prolific formation of low mass
galaxies in the standard Cold Dark Matter scenario. Vedel, Hellsten \&
Sommer-Larson (1994) used simulations of the collapse of an overdense,
rigidly-rotating sphere to argue that the central regions of galaxies
as massive as the Milky Way can be influenced by photoionization
effects. This qualitative conclusion, however, was based on a single
simulation and therefore cannot be used to derive results of general
applicability. On the other hand, Steinmetz (1995), Quinn, Katz \&
Efstathiou (1996), Thoul \& Weinberg (1996) and Weinberg, Hernquist \&
Katz (1996) have recently concluded that in systems with circular
velocities larger than about $\sim 50$ km s$^{-1}$ the effects of a UV
background would be rather small. Their conclusions are based on the
results of 3-$D$ and 1-$D$ simulations of the formation of low mass
systems in the CDM scenario.

This apparent discrepancy could be due to a number of reasons. (i) The
poor spatial resolution of Cen \& Ostriker may have artificially
enhanced the effects of the UV background (poor numerical resolution
mimics heating and leads to a reduction in the efficiency of
cooling). This seems likely given the results of Weinberg et al., who
demonstrate that insufficient numerical resolution can artificially
suppress the formation of low mass objects. (ii) The artificial
initial conditions of Vedel et al. complicates the interpretation of
their results. (iii) Steinmetz, Quinn et al., and Weinberg et
al. studied the formation of low-mass halos at $z=3$, $z=2.4$, and
$z=2$, respectively, and therefore their conclusions are not directly
applicable to present-day galaxies. (iv) The spherical symmetry
assumed by the 1D treatment of Thoul \& Weinberg hampers the
comparison between their results and those of the previous studies.

We present in this paper 3D N-body/gasdynamical simulations of the
formation of galaxies in the standard CDM scenario designed to address
these issues. In particular, we focus on how the presence of a UV
background affects the amount of gas that can cool in dark matter
halos, as well as on the rotational properties of gaseous disks formed
in these systems.  We improve on previous studies by evolving up to
the present ($z=0$) systems with circular velocities between $80$ and
$200$ km s$^{-1}$. Our mass and spatial resolution also represent a
significant improvement over previous studies ($\sim 5000$ gas
particles per system compared with $\lsim 1000$ of Quinn et al.). As in
the studies mentioned above, we neglect the effects of star formation
in this simplified approach.

The plan of this paper is as follows. In \S 2 we describe the
numerical code and the initial conditions of the simulations. Section
3 describes the results of the numerical experiments. The time
evolution of the systems for different choices of the spectrum and
redshift dependence of the photoionizing background is presented in
\S3.1; \S 3.2 discusses the amount of mass that can cool in each of
these systems as a function of redshift and of the strength of the
radiation field. In \S3.3 we analyze the rotational properties of the
disks. Section 4 compares these results with previous works and
discusses their implications for hierarchical models of galaxy
formation. Section 5 contains a brief summary of our main
conclusions. 

\section{The Numerical Experiments} \label{numexp}

\subsection{The Code}

The numerical code we use is GRAPESPH, an implementation of a hybrid
N-body/hydrodynamical code for the special purpose hardware GRAPE
(Sugimoto et al. 1990). This code combines a direct-summation N-body
integrator with the Smooth Particle Hydrodynamics (SPH) approach to
numerical hydrodynamics (Monaghan 1992). It is described in detail in
Steinmetz (1996). The version of the code we use here is especially
adapted to evolve a mixture of collisionless (dark matter) and
collisional (gas) fluids. It is fully Lagrangian, three-dimensional,
and highly adaptive in time and space by means of individual smoothing
lengths and individual timesteps. The physical processes included in
this version of the code include self-gravity, pressure gradients,
hydrodynamical shocks, radiative and Compton cooling and, optionally,
the photoheating of a UV background with a specified spectrum. The
code follows self-consistently the non-equilibrium time evolution of
the six baryonic species (H, H$^{+}$, He, He$^{+}$, He$^{++}$, and
$e^-$). We assume that the gas remains optically thin throughout the
calculation and that the background radiation is uniform in space;
ie. we neglect any corrections due to radiative transfer effects.

One important feature of the code we use here is that it uses a
``shearfree'' formulation of the artificial viscosity in the SPH
equations of motion. This is important to prevent artificial viscous
angular momentum transport in gaseous disks. A full description of the
shearfree viscosity implementation, as well as of tests relevant to
the issues we discuss in this paper, are presented in the Appendix.

\subsection{The Initial Conditions}

We simulate the evolution of 8 different systems with circular speeds
in the range $80$ \kms$ < V_c < 200$ \kms. These systems are selected
from a large cosmological simulation of a standard $\Omega=1$ CDM
universe carried out with a P$^3$M code (Efstathiou et al. 1985). This
simulation follows the evolution of $64^3$ particles in a periodic box
of $30$ Mpc on a side. (Here and throughout this paper we assume a
value of $H_0=50$ km s$^{-1}$ Mpc$^{-1}$ for the Hubble constant and
for all physical quantities that depend on it.) The present time
($z=0$) is identified by requiring that the linear {\it rms}
fluctuations in a sphere of radius $16$ Mpc equals $\sigma_8 = 1/b =
0.63$ ($b$ is the usual ``bias'' parameter). The eight systems are
selected from a list of clumps compiled using a friends-of-friends
algorithm with linking parameter set to $10\%$ of the mean
interparticle separation.

In order to span a large range in mass but to have as well some
indication of the ``cosmic'' scatter at a given mass we choose the
eight systems in two mass bins; four low-mass systems with $V_c$
between $80$ and $100$ km s$^{-1}$ and four massive systems with $V_c$
between $160$ and $200$ km s$^{-1}$.  The particles of each system are
then traced back to the initial conditions, where a box containing all
of them is drawn. The size of the box is $4.5$ and $7.5$ comoving Mpc
on a side for the low- and high-mass systems, respectively. We load
this high-resolution box with $40^3$ particles on a cubic grid and
perturb them with the same long waves as in the original P$^3$M
simulation, plus some shorter waves up to the Nyquist frequency of the
new particle grid. The outer regions are coarse-sampled with a few
thousand particles of radially increasing mass in order to reproduce
the tidal effects due to distant material.  This procedure ensures the
formation of a clump that is indistinguishable from the one selected
in the P$^3$M run except for the improved spatial and mass
resolution. Further details can be found in Navarro \& White (1994).

The gas component is included by laying the same number of gas
particles on top of the dark matter particles in the high-resolution
box. They are given their same velocities, and are assigned masses
assuming a mean baryon density of $5\%$ ($\Omega_b=0.05$). The mass
per gas particle is $4.9 \times 10^6 M_{\odot}$ and $2.3 \times 10^7
M_{\odot}$ in the low- and high-mass systems, respectively.  We adopt
a Plummer gravitational softening of $2.5$ ($5$) kpc for the dark
matter and of $1.25$ ($2.5$) kpc for the gas particles in the low
(high) mass systems. All runs are started at $z=21$. The number of
timesteps varies between $50,000$ and $1,500$ for particles in the
densest regions and in the boundary, respectively.  We have also
carried out a series of low-resolution runs, following exactly the
same procedure as before but reducing the number of particles in the
high-resolution box by a factor of six. This provides a direct test of
the importance of numerical resolution effects.

\subsection{The Photoionizing Background}

Observations of the observed deficit of Ly$\alpha$ forest lines near
quasars (the ``proximity effect'') can be used to put some constraints
on the UV background at high redshifts. A compilation of ground-based
data for high-redshift quasars plus data from the HST FOS Key Project
indicates that the specific intensity of the UV background at the
Lyman limit, $J_{\nu}(\nu_L)$, seems to increase with redshift to
reach $J_{\nu}(\nu_L) \approx 10^{-21} {\rm erg \ s}^{-1} {\rm
cm}^{-2} {\rm sr}^{-1} {\rm Hz}^{-1}$ at $z \sim 2$-$3$. There is
preliminary evidence of $J_{\nu}(\nu_L)$ leveling off or even
declining at higher redshifts. (For a review, see Bechtold 1995.)
Based on these results, we have chosen to model the time evolution of
the ambient UV radiation by

$$ J_{\nu}(z)={J_{21}(z) \times 10^{-21}
\biggl({\nu_L \over \nu}\biggr)^{\alpha}
{\rm erg \ s}^{-1} {\rm cm}^{-2} {\rm sr}^{-1} {\rm Hz}^{-1}}
\eqno(1)
$$

We use $\nu_L$ to denote the Lyman limit and we parameterize
$J_{21}(z)$ as in Vedel et al. (1994),

$$
J_{21}(z) = {J_{21}^0 \over 1 + \bigl({5 / 1+z}\bigr)^4}. \eqno(2)
$$

Most of our runs use $J_{21}^0=1$ and $\alpha=5$, but we have also
explored, in two runs, an extreme version of the photoionizing
spectrum, assuming $J_{21}^0=10$ and $\alpha=1$. The time evolution of
$J_{21}$ in these two cases is compared with observational constraints
in Figure 1.  Our choice is likely to overestimate the effect of the
UV radiation field at $z>4$.  However, we shall see below that the
density of collapsing clumps at $z>4$ is so high that even the more
energetic background has almost negligible effects on the cooling
efficiency.  The true evolution of the UV radiation field is likely to
fall between our two models at lower $z$.

Table 1 gives a complete summary of the relevant parameters of each
run and introduces useful notation. Runs $1$-$4$ refer to low-mass
systems, runs $5$-$8$ to high-mass systems. We have labeled each run
that includes a UV background with ``$J$''; the extreme UV background
runs have an extra ``$a$'' index. Finally, runs preceded by ``$L$''
refer to low-resolution models, where the number of particles has been
artificially lowered by a factor of 6.  The columns $M_{200}$,
$r_{200}$, and $V_{200}$ give the mass, radius and circular velocity
of a sphere within which the mean overdensity of the system is 200 at
$z=0$. We shall refer to $r_{200}$ as the ``virial'' radius. $N_{gas}$
and $N_{DM}$ are the numbers of gas and dark matter particles within
that sphere. $M_{DM}$ and $M_{gas}$ are the total mass in each
component within $r_{200}$. The mass of the central gaseous disk at
$z=0$ is listed under $M_{disk}$.

\section{Results}\label{results}
\subsection{Time evolution}\label{evol}

The general evolution in runs that do not include a UV background
(runs $1$-$8$) is very similar to that described in Navarro
\& White (1994) and Navarro et al. (1995a). 
Here we briefly describe the salient aspects of the typical formation
process of these systems, and refer the reader to those papers for a
more detailed discussion.  Clumps of small mass collapse at high
redshift and merge progressively to form more massive systems. The
largest progenitor has about half the final mass at $z=1$-$1.5$. The
gas follows the dark matter as it expands, turns around, and collapses
in the smallest resolved clumps. The energy gained during the collapse
is converted into heat by shocks and quickly radiated away. The gas
then sinks to the center of the halos, where it settles into
centrifugally supported disks. More massive disks are formed by
mergers of smaller ones plus the smooth accretion of gas particles in
smaller, unresolved clumps.

Figure 2 shows one of the systems at different evolutionary stages.
Note that as early as $z=6$ a significant fraction of the mass of the
systems is already in collapsed clumps. At this time, the mass of the
largest progenitor is about a tenth of the final mass. The last major
merger seems to occur at $z \gsim 1$, and only a small number of
low-mass satellites are accreted after $z\sim 0.5$.  The inclusion of
a UV background prevents some gas from cooling at late times, although
this effect is difficult to appreciate in Figure 2. The disk at $z=0$
appears, however, much more concentrated in this case. Note the
presence of a few satellites orbiting around the central disk, some of
them stripped of their dark matter halos. These satellites contain
only a small fraction of the gas mass of the system, most of which
resides in the central disk.  We discuss these effects in more detail
in the following subsections.

\subsection{The mass of cooled gas}

\subsubsection{Runs without a UV background}

Figure 3 shows the evolution of the gas component in the
density-temperature plane for the system shown in Figure 2. The
horizontal lines in these panels correspond to the mean baryon density
at each redshift. The meandering curve separates the regions where the
cooling time, $T_{cool}$, is shorter than the current age of the
universe, $t(z)$. Particles expanding or contracting adiabatically
move along lines of slope $(\gamma-1)^{-1}=1.5$ in this plot. This is
the behavior seen for underdense regions at $z=9$ and $z=6$. At the
same redshifts, particles in overdense regions are distributed almost
horizontally; these are particles in collapsing clumps that are being
heated by shocks. Since even strong shocks cannot raise the density by
more than a factor of few but there is no bound on the jump in
temperature, shock-heated particles appear to stretch almost
horizontally in this plot.

In the absence of cooling, gas particles would be heated to the virial
temperature of the halo they belong to; $T_{vir}=(1/2) \mu m_p V_c^2/k
\approx 36 (V_c/$km s$^{-1})^2$ K (assuming $\mu \approx 0.6$ as
appropriate for a plasma of primordial composition). This corresponds
to $\sim 10^6$ K for a galaxy with circular speed of $\sim 200$
\kms. However, once the temperature of a shock-heated particle reaches
about $10^4$ K, it crosses the boundary drawn by the cooling curve;
cooling becomes very efficient and the particle is forced to remain at
that temperature. Lost its pressure support, the gas collapses to the
center of the halo, where it forms centrifugally supported disks such
as that seen in Figure 2.  At later times, as the density of the
universe drops, particles are seen to ``tunnel'' past the cooling
boundary to the virial temperature. Because for a primordial plasma
cooling is not very efficient at $T \sim 10^6$ K, particles that can
reach these high temperatures remain there relatively undisturbed,
forming a high-pressure, low density environment that surrounds the
cold and dense disks.

As is clear from Figure 3, the amount of gas that cools in a galactic
halo depends strongly on the redshift at which clumps with potential
wells deeper than $T_{vir} = 10^4$ K collapse. If much of this
collapse happens at $z \gsim 1$ the fraction of cooled gas is expected
to be high. Gas accreted later finds it easier to tunnel past the
$10^4$ K minimum in the cooling curve to high-temperatures, where
cooling times are long. Gas in this hot-phase can remain in the halo
for more than a Hubble time and would not participate in the formation
of the main body of a galaxy. (We note that this conclusion depends
sensitively on the primordial composition assumed. Cooling rates can
be substantially boosted by metals at $T \sim 10^6$ K; by more than an
order of magnitude for solar composition. Therefore, a significant
fraction of the gas in the hot atmosphere would be able to cool if it
were sufficiently enriched by metals.)

Cooled gas sinks almost isothermally at $10^4$ K to the center of dark
matter halos, increasing its density until the collapse is stopped by
the formation of a centrifugally supported disk. Material in these
disks is usually at $\rho > 10^{5} M_{\odot}/$kpc$^3$. Since at these
densities cooling dominates at essentially all $z$ for $T> 10^4$ K, we
shall adopt hereafter this value to distinguish between material in
dense, centrifugally supported disks from infalling or
pressure-supported gas. (For reference, $10^5 M_{\odot}/$kpc$^3$ is
about the baryonic density in the solar neighborhood, see Table 1-1 in
Binney \& Tremaine 1987.)  At $z=0$ most of the gas ($88 \%$) is in
the central disk (Table 1). Note that this is a lower limit to the
amount of cooled gas in the system, since it does not include the cold
gas in small satellites orbiting around the central disk. Limited
numerical resolution will also tend to decrease the total amount of
cooled gas by preventing the collapse of low-mass clumps at high
redshift, just when cooling is most effective. The effects of
resolution can be assessed by reducing the number of particles in a
run. Table 1 shows that, at $z=0$, the effects of our limited
resolution seem to be small. The fraction of gas in the central disk
decreases from $88 \%$ to just $87 \%$ when the number of particles is
reduced by a factor of 6 (compare runs $5$ and $L5$ in Table 1). The
effects of poorer numerical resolution are, however, more readily
noticeable at higher redshift (see \S3.23 below and Figure 5).

\subsubsection{The effects of a UV background}

Figure 4 shows the $\rho$-$T$ evolution of the gas once a UV
background is included ($J_{21}^0=1.0$, $\alpha=5$, run $J5$). The
horizontal line and the curve on the right are as in Figure 3. The
cooling curve has been substantially modified relative to that in
Figure 3 because hydrogen and helium are highly ionized and,
therefore, cooling by recombination and collisional excitation are
strongly suppressed.  The curve on the left indicates where the
photoheating timescale equals the age of the universe, and the
nearly-vertical line indicates the equilibrium temperature (ie. where
the cooling and heating rates balance out).

One of the main differences between runs with and without UV
background is that in the former, the gas is photoheated to about
$10^4$ K before it is accreted into resolved non-linear clumps. (Note
that at $z=9.5$ that the gas is slightly cooler than the heating
timescale would predict. This is due to the effect of Compton cooling,
which is especially important at high $z$). At $z=9$ and $z=6$ the gas
densities are still high enough for cooling by collisional processes
($\propto \rho^2$) to dominate over heating by photoionization
($\propto \rho$). Cooling is quite efficient and practically all the
gas that gets accreted into systems with virial temperatures $T>10^4$
K cools almost immediately to the equilibrium temperature.  By $z=2$
the reduced cooling capabilities of low density gas become more
evident; gas needs to be about one order of magnitude denser than the
mean in order to cool efficiently, and denser still at lower
redshifts. As a result, a significant fraction of the gas accreted
into clumps at $z< 2$ can reach the virial temperature of the final
halo ($\sim 10^6$ K) at relatively low overdensities. Because of its
long cooling timescale, much of this gas remains hot and diffuse until
$z=0$. As we noted in \S3.2.1, cooling times at $T \sim 10^6$ K can be
substantially shortened by the inclusion of metals. Introducing metals
would thus counter the effects of photoheating, resulting in higher
cooling efficiencies similar to those achieved in the absence of a UV
radiation field.

This reduction in the cooling efficiency can be clearly seen by
comparing the mass of the central disk in runs with and without UV
background. The central disk mass is reduced by about $30 \%$ by the
inclusion of our fiducial photoionizing background. The reduction is
more pronounced for more energetic radiation fields. In runs $J4a$ and
$J5a$ ($J_{21}^0=10$, $\alpha=1$) the mass of the central disk is
reduced by almost one-half. The disk masses at $z=0$ do not seem to be
significantly affected by numerical resolution; on average, the
low-resolution runs that include a UV background have the same disk
mass as do the high-resolution models. We shall see below that the
situation is different at higher redshifts, when the effects of
numerical resolution become more pronounced.

\subsubsection{The time evolution of cooled gas}

The time evolution of the gas that cools is shown in Figure 5.  Here
we plot, as a function of $z$, the mass of gas in dense disks (defined
by $\rho > 10^5 M_{\odot}/$kpc$^3$, as discussed above), expressed as
a fraction of the final gas mass of each system. Each curve is an
average over all simulations with the same UV parameters. This figure
confirms quantitatively the qualitative picture indicated by Figures 3
and 4. At $z \sim 10$ there is little difference in the fraction of
cooled gas regardless of the presence of a UV background. The rate at
which gas cools subsequently, however, depends strongly on the
intensity of the background. For our fiducial values of $J_{21}^0=1$
and $\alpha=5$, the influence of photoionization on the mass of cooled
gas is negligible until $z \sim 3$. Later evolution is affected
significantly more, but still the overall effect is mild and reduces
the total fraction of cooled gas from $90 \%$ to about $75 \%$. A
harder background ($J_{21}^0=10$, $\alpha=1$) has a more noticeable
effect, reducing the cooled fraction to about $55 \%$.

Figure 5 also illustrates the effects of numerical resolution. With
fewer particles the collapse of low mass clumps at high-$z$ is not
well resolved, causing a severe underestimate of the total amount of
gas that can cool. As clumps get better resolved with time, the
cooling rate accelerates in the low-resolution runs (symbols connected
with dotted lines) and at $z=0$ the total fraction of cooled gas
converges to the values found in the high-resolution runs. The effects
of poor numerical resolution are, therefore, to delay the cooling and
aggregation of gas into dense clumps. This may have undesirable
consequences, for we shall see below that the rotational properties of
the gas component depend sensitively on the timing and mode in which
gas is accreted into the system.

We cannot rule out the possibility that even our high-resolution runs
are affected by numerical limitations, especially at high
redshift. From the discussion of Figure 3, we may expect that in the
absence of a UV background the rate at which gas cools should in
principle be limited only by the rate at which clumps with virial
temperatures larger than $10^4$ K are formed. We can use the
Press-Schechter theory to compute these rates. The mass fraction in
clumps with masses larger than $M$ at redshift $z$, subject to the
constraint that it will form part of a halo of mass $M_0$ at $z=0$, is
given by (Bower 1991, Bond et al. 1991, Lacey \& Cole 1993)

$$
f(>M,z| M_0)= {\rm erfc} \biggl( { 1.69 \times z \over
\sqrt{2(\Delta^2(M) - \Delta^2(M_0))}}\biggr). \eqno(3)
$$

Here $\Delta^2(M)$ is the variance of the linear power spectrum at
$z=0$ smoothed with a top-hat filter of radius $r_0$ enclosing a mass
$M=(4\pi/3) \rho_0 r_0^3$. ($\rho_0$ is the mean density of the
universe at $z=0$.) The corresponding circular velocity is given by
$V_c=1.677 (1+z)^{1/2} H_0 r_0$.  The cumulative mass in clumps with
$V_c > 16.6$ \kms, the circular velocity corresponding to a virial
temperature of $10^4$ K, is shown in Figure 5 with dashed lines. The
upper and lower curves correspond to the constraint that at $z=0$ all
the mass should be in clumps with $V_c=200$ and $80$ \kms,
respectively. Note that the curves depend very weakly on circular
velocity in the range probed by our simulations. If the cooled gas
mass is determined principally by the rate at which gas is accreted
into clumps with $T_{vir}>10^4$ K we would expect at most a weak
dependence on circular velocity. We shall see below that this is
indeed the case.

The comparison between high- and low-resolution runs show that, as the
numerical resolution improves, the curves corresponding to runs
without a UV background approach the dashed lines. The vertical offset
between these curves can then be interpreted as a measure of the
effect of numerical limitations on the mass of cooled gas. We can
reasonably expect that, were the resolution improved further, the
curves with solid symbols would move closer to the dashed
lines. Despite this potential deficiency, our simulations confirm that
the cooling efficiency is indeed very high in the absence of a heating
mechanism, as discussed in \S1. They also show that limited numerical
resolution always tends to reduce the efficiency of cooling. The
curves in Figure 5 and the amounts quoted in Table 1 should therefore
be viewed as lower limits on the mass of cooled gas, rather than as
firm determinations. Estimates of the amount of cooled gas seem to be
particularly vulnerable to numerical resolution at high redshift, when
the typically smaller masses and sizes of halos make them more
difficult to resolve adequately in a given simulation.

\subsubsection{The mass of the central disk}

Most of the gas that cools is rapidly accreted into the central disk,
in good agreement with the results of Navarro \& White (1994) and
Navarro  et al. (1995a). This is shown in Figure 6, where we plot the
total mass of the central disk as a function of $V_{200}$, the
circular velocity at the virial radius $r_{200}$. By definition, the
total mass within $r_{200}$ is $M_{200}=200 (4\pi/3) \rho_0 r_{200}^3$
and $V_{200}^2=GM_{200}/r_{200}$. The solid and dotted lines
represent, respectively, the dark matter and the gas mass within
$r_{200}$ expected from our choice of baryon density;
$M_{DM}=(1-\Omega_b) M_{200} \propto V_{200}^3$, and $M_{gas}=\Omega_b
M_{200} \propto V_{200}^3$. Open circles are used for the runs without
UV background, filled squares for the ``$J$'' runs, and starred symbols
for the ``$Ja$'' runs.

This figure shows clearly that a large fraction of the gas mass in the
system ends up in the central disk.  The presence of a UV background
can reduce the disk mass somewhat. The magnitude of the effect seems
to be almost independent of the circular velocity of the system,
affecting low mass halos almost as much as more massive systems. Tight
correlations are observed between the circular velocity of the halo
and the mass of the disk. Fits of the form $M_{disk}
\propto V_{200}^n$ have very small scatter; $0.035$ in log $M_{disk}$
($n=2.75$) for runs without background and $0.057$ in log $M_{disk}$
(n=2.45) if our fiducial background is included. This small scatter is
consistent with the small dispersion in the observed Tully-Fisher
relation $\sim 0.16$ in log $L$ (Willick et al. 1995). We caution,
however, against overinterpreting this result. The estimate of the
scatter is based on eight simulations only. Also, the comparison with
the Tully-Fisher relation assumes that disk mass is directly
proportional to luminosity and that the circular velocity of the halo
is identical to the rotation speed of the disk. Both of these
assumptions may be incorrect (see Navarro, Frenk \& White 1996 for a
discussion). Nevertheless, the scatter found seems to be significantly
smaller than predicted by the analytic arguments of Eisenstein \& Loeb
(1996). A larger set of simulations is needed in order to clarify the
reasons for this discrepancy.

\subsection{The angular momentum of gaseous disks}

Figure 7 shows the specific angular momentum of the dark matter halos
and of their central gaseous disks at $z=0$, as a function of
mass. The boxes indicate the loci corresponding to spiral and
elliptical galaxies, as compiled by Fall (1983). The symbols are as in
Figure 6. This figure shows clearly that the spins of gaseous disks
formed in the absence of a UV background (open circles) are about an
order of magnitude lower than that of their surrounding halos. This is
a direct consequence of the formation process of the disks (Navarro et
al.  1995a). Most of the disk mass is assembled through mergers
between systems whose own gas component had previously collapsed to
form centrally concentrated disks. During these mergers, and because
of the spatial segregation between gas and dark matter, the gas
component transfers most of their orbital angular momentum to the
surrounding halos (Frenk et al. 1985, Barnes 1988, Quinn \& Zurek
1988).

We illustrate this process in Figure 8, where we show the time
evolution of the halo (dotted lines) and central disk (solid lines) of
two different systems in the $J/M$ vs $M$ plane. The open circles
correspond to $z=5$, when each system had only a few hundred
particles. Note that at this redshift the spins of halos and disks are
similar. This is most likely an artifact of poor numerical resolution;
at high redshift the collapse of gas in a system made of only a few
hundred particles is artificially smooth and allows the gas to retain
much of its angular momentum. As the mass of the system increases the
angular momentum of the halo grows, following approximately the
relation $J \propto M^{5/3} (1+z)^{-1/2}$ expected if halos at all
times had roughly the same value of the dimensionless spin parameter
$\lambda=J |E|^{1/2}/GM^{5/2}\approx 0.05$ (Barnes \& Efstathiou 1987,
Steinmetz \& Bartelmann 1995). (Here $E$ is the total binding energy
of the system.) The spin of the central disk does not follow the same
relation, decreasing steadily as more of its mass is accreted in the
form of mergers between systems that are well resolved numerically.
We stress that the loss of angular momentum seems to be a sensitive
function of numerical resolution. In the low-resolution runs the final
spins of the disks are systematically higher (by up to a factor of
$\sim 3$). Two factors contribute to this increase. (i) The collapse
of a system of only a few hundred particles is artificially smooth,
and (ii) the cooling and accretion of gas into the final disk are
delayed (Figure 5). Both of these effects conspire to allow the gas
component to retain more of its angular momentum in low resolution
runs. We shall come back to this issue in \S4.

Figure 7 shows that disks formed under the influence of a UV
background suffer from the same angular momentum problem. Indeed, the
spins of disks formed in the ``$J$'' runs (solid squares) and in the
``$Ja$'' runs (starred symbols) are on average even lower than those
formed without a UV radiation field, exacerbating the discrepancy
between observed spirals and the gaseous disks formed in the
simulations. The reason why this problem is aggravated by the UV
radiation field can be seen in Figure 9. This figure shows, for the
disk in run $5$, the correlation between the angular momentum of a gas
particle at $z=0$ and the time at which that gas element is accreted
into the disk. Particles accreted later clearly have larger angular
momenta. This implies that disks grow from the inside-out, as mass
accreted later settles at progressively larger radii from the center
of the disk. As discussed in \S 3.2.3 (Figure 5), photoheating
inhibits preferentially the late accretion of gas. This deprives the
disk from its richest source of angular momentum and accentuates the
discrepancy with observed spirals.

A different way to show that the spins of the gaseous disks are
deficient compared with spirals is to compare their sizes with the
luminous radii of spiral disks. As noted by Fall \& Efstathiou (1980),
if during the formation of the disk the gas conserves the same angular
momentum as that of its surrounding halo, the size of the disks formed
would be given approximately by $\sim \lambda r_h$, where $r_h$ is the
half-mass radius of the dark halo and $\lambda \sim 0.05$ is the
typical rotation parameter of the system. Figure 10 shows that this
identification would predict a disk half-mass radius of $\sim 6$ kpc
for a circular speed of $\sim 200$ \kms (dotted line), in good
agreement with observations of galaxies like the Milky Way. However,
because the angular momentum content of gaseous disks is well below
that of the halos (Figure 7) the half-mass radii of the disks is much
smaller than those of spiral galaxies.

In summary, the assembly of gaseous disks through mergers leads to the
formation of systems that are too concentrated or, alternatively,
whose spins are too low to be consistent with observations of spiral
galaxies. Photoionization seems only to exacerbate this problem.

\section{Discussion}

The simulations presented above indicate that the presence of a
photoionizing background can have a significant influence on the
amount of gas that can cool in galactic halos. For halos with circular
velocities between $80$ and $200$ km s$^{-1}$ the cooled gas fraction
is $\sim 90\%$ in absence of a UV radiation field, and is reduced by
up to $\sim 50\%$ when an extremely energetic background is included
($J_{21}^0=10$, $\alpha=1$). The reduction seems to be essentially
independent of halo circular velocity in the range considered.  Most
of the reduction is due to the low cooling efficiency of gas that gets
accreted late into systems with virial temperatures above $\sim 10^4$
K.  At high redshifts the mass of cooled gas is independent of the UV
background.

These results disagree somewhat with those of Thoul \& Weinberg
(1996), who concluded that above $V_c \sim 75$ \kms the influence of
photoionization would be negligible. The main discrepancy between
their and our results seems to be in the cooled gas fraction obtained
in {\it absence} of photoheating. These authors find that the cooled
fraction is $\lsim 50 \%$ in systems with $V_c>75$ \kms, and that it
tends to decrease even further for larger $V_c$. This is significantly
lower than the average $90 \%$ of gas that cools in our simulations,
especially taking into account that this is likely to be a lower limit
(see \S 3.2.3). The reason for the discrepancy may be the idealized
1-$D$ treatment of Thoul \& Weinberg. Mass is accreted in this model
by radial shells that successively turn around and collapse. The
characteristic density of each shell then decreases with time. This
differs markedly from the behavior illustrated in Figure 2, where we
see that even mass that gets accreted ``late'' into the system may
collapse at higher redshift into a different halo. This effect can
conceivably lead to systematic underestimates in the cooling
efficiency of gas in 1-$D$ simulations such as those of Thoul
\& Weinberg.

Quinn et al. (1996) find a cooled gas fraction of about $50\%$ in
their simulations of a system with a circular speed of $46$ \kms
system at $z=2.4$ in the absence of photoionization. Their simulations
use numerical techniques similar to ours and have only slightly worse
resolution, so we would expect our results to compare favorably. Based
on the results presented in Figure 5 we would have predicted a higher
cooled gas fraction in their system, although this expectation is
based on an uncertain extrapolation of our results to systems of lower
mass and higher collapse redshifts that the ones we have
considered. These authors also find that including the same extreme
photoionizing background as ours ($J_{21}^0=10$, $\alpha=1$) can
reduce the cooled gas fraction by more than $40 \%$, in good agreement
with our results. With the caveat that Quinn et al. consider systems
of different mass than ours and stop their simulations at $z=2.4$, we
conclude that our and their results are in reasonably good agreement.

Our results are also in qualitative agreement with those of Weinberg
et al. (1996), who found that photoionization does not have a
significant effect on the mass of cooled gas at $z=2$, as long as
numerical resolution effects are properly taken into account. The
ionizing background chosen by these authors in their
``high-resolution'' simulation is weaker than our fiducial choice, so
we would also predict no significant effects between runs with and
without photoheating. This is especially true at $z=2$, when these
authors stopped their simulation.

Taken together, all these results imply that the collapse and cooling
of gas in galactic halos can be at most moderately affected by the
inclusion of a photoionizing background. In systems with circular
speeds as low as $50$-$80$ \kms at least $50\%$ of the gas would be
able to cool even if an extremely hard UV background were
present. The magnitude of this effect is much smaller than needed in
order to reconcile the large number of low-mass halos predicted by
hierarchical clustering models with the relative paucity of dwarf
galaxies in the local universe. According to the analysis of Cole et
al.  (1994, see their Figure 2b), the fraction of gas that can cool
and form stars should be reduced to less than about $10\%$ in systems
with $V_c=50$ \kms and to $\lsim 25 \%$ in systems with $V_c=100$ \kms
in order to reconcile the shape of the galaxy luminosity function with
the halo mass function in the Cold Dark Matter scenario.

Photoionization effects also prove ineffective at alleviating the
angular momentum problem of gaseous disks assembled in hierarchical
scenarios. By reducing preferentially the late accretion of mass into
disks (Figure 5) photoheating reduces further the already low spin of
gaseous disks formed through mergers.  It is important to prove that
this result is not the consequence of some subtle numerical effect,
especially because the disks reported in the simulations of Katz \&
Gunn (1991), Vedel et al, (1994), Steinmetz \& M\"uller (1995), and
Evrard, Summers \& Davis (1994) allegedly have sizes and spins
consistent with observed spirals even though they used similar
numerical techniques to the ones we used here.

To reason for this apparent discrepancy is that the transport of
angular momentum from the gas to the dark matter during mergers is a
sensitive function of the degree of spatial segregation between gas
and dark matter in the merging protogalaxies. Therefore, the final
spin of the gaseous disk depends on (i) the importance of mergers in
the assembly of the mass of the system, and on (ii) the gas fraction
that had collapsed to the center of the protogalaxies previous to each
merger event. The initial conditions adopted by Katz \& Gunn (1991),
Vedel et al (1994), and Steinmetz \& M\"uller (1995) (overdense,
homogeneous spheres endowed initially with solid-body rotation and
perturbed with small scale density fluctuations) emphasize the role of
coherent collapse during the formation of the system and minimize the
role of mergers. The disagreement with the results of Evrard et
al. (1994) is, on the other hand, likely to be numerical in
origin. Most of the systems analyzed by Evrard et al. (1994) have only
a few hundred particles.  As noted when discussing Figure 8, the
collapse of systems of only a few hundred particles is artificially
smooth and allows the gas to retain much of its angular momentum. Only
when the proper initial conditions are used and the numerical
resolution is adequate does the discrepancy between the spins of
gaseous disks and those of spiral galaxies become apparent.

Another numerical effect which could in principle lead to artificially
small disk sizes is the use of an artificial viscosity term in the SPH
equations of motion. Introduced in order to model shock waves, it is
usually implemented so as to damp the relative (approaching) motion of
any two neighboring gas particles. This implies that in a gaseous disk
differential rotation is seen by neighboring particles at slightly
different radii as convergent motions and consequently damped. The
result is an artificial transfer of angular momentum to the outer
regions of the disk. Mass is channelled inwards in order to conserve
angular momentum, leading to the formation of disks with artificially
high central concentrations. This is a concern not only because it
could affect adversely the spatial structure of the disks, but also
because it could conceivably lead to enhanced losses of angular
momentum during mergers (disks that are artificially small will take
longer to merge and may lose more angular momentum in the
process). These concerns prompted us to use a shear-corrected
implementation of the artificial viscosity. (See Appendix A for a full
description of the artificial viscosity used and the tests performed.)
This correction essentially eliminates the viscous angular momentum
transport in disks. We conclude that the angular momentum deficit of
gaseous disks in our simulations are of physical, rather than
numerical, origin.

As discussed in \S 2.3, we expect the strength of the UV background
adopted in the simulations to be reasonably consistent with
observations at intermediate and high redshifts. At very high ($z>4$)
and very low ($z\la 0.5$) redshifts, however, the observational
constraints are quite poor. We believe, however, that our results are
fairly insensitive to most reasonable variations of the ionization
history at these epochs.  At $z>4$ the density of the universe is so
high that the efficiency of cooling is hardly affected by the UV
background.  At lower redshifts the situation is also unlikely to
change. Most of the gas that is prevented from cooling by the presence
of the UV radiation field ends up in a low density, hot halo where the
cooling times are longer than a Hubble time (see Figures 3 and 4). At
these temperatures the main cooling mechanism is bremsstrahlung and,
consequently, independent of the UV field.  We do not expect our
conclusions to change dramatically for any plausible variations of the
UV background strength.


We conclude that photoionization cannot provide the heating mechanism
desired to bring hierarchical galaxy formation models into agreement
with observations. A different heating process such as feedback
effects related to star formation and evolution must therefore be
invoked. Parameterizations of this feedback mechanism by White \&
Frenk (1991), Cole et al. (1994) and Kauffmann et al. (1994) show that
this procedure can in principle work. Unfortunately, because of our
primitive understanding of the process of star formation, such
mechanism can only be introduced in our numerical schemes only in the
form of numerical ``recipes'' containing a number of free
parameters. Only a painfully exhaustive search of the parameter space
seems likely to provide new insights into the complex process of the
formation of galaxies in hierarchical clustering models.

\section{Summary}

We have presented numerical simulations designed to investigate the
influence of a photoionizing background on the formation of disk
galaxies in hierarchically clustering universes. These simulations
span the mass range of typical galaxies ($80$ \kms $< V_c < 200$ \kms),
follow the evolution of systems until $z=0$, and have better numerical
resolution than earlier studies. The main results of our simulations
can be summarized as follows. 

\items{1)
In the absence of a UV photoionizing background most of the gas ($\sim
90 \%$) in galactic halos cools and assembles into cold, dense,
centrifugally supported disks. The presence of a UV background can
reduce the amount of gas that cools, but the effect seems to be mild
($\sim 15 \%$) for our fiducial choice of background ($J_{21}^0=1,
\alpha=5$). Cooled gas masses can be reduced by up to a factor of
$\sim 2$ for an extreme version of the background ($J_{21}^0=10,
\alpha=1$). The magnitude of this effect is insufficient to reconcile
the shape of the galaxy luminosity function with the mass function of
dark matter halos. }

\items {2) 
The influence of the UV background seems to be largely independent of
circular velocity, at least in the range, $80$ \kms $ < V_c < 200$
\kms, probed by our simulations. We interpret this as implying that
the effect of the background depends principally on the rate at which
the mass that makes up the final system gets accreted into clumps with
virial temperatures higher than $\sim 10^4$ K collapse. This is
essentially independent of $V_c$ in the range considered.}

\items{3)
The UV radiation field affects primarily the late infall of gas and
has negligible influence on the amount of cooled gas at high redshift
($z>5$).  Since disks grow inside-out, accreting the outer,
high-angular momentum regions at late times, the ionizing radiation
tends to reduce the total spins and sizes of gaseous disks assembled
at the center of dark matter halos. This exacerbates the angular
momentum problem of gaseous disks assembled in hierarchical clustering
scenarios.}

We conclude that a different heating mechanism needs to be invoked in
order to bring hierarchical models of galaxy formation into agreement
with observations. The energy feedback of evolving stars and
supernovae seems the most obvious and attractive alternative, implying
that an improved understanding of galaxy evolution is intimately
related to progress in our understanding of star formation and of the
interaction between evolving stars and the interstellar medium. The
effects of star formation are thus crucial ingredients of any
successful model of galaxy formation.

\acknowledgments

We would like to thank the hospitality of the Institute for
Theoretical Physics of the University of California at Santa Barbara,
where some of the work presented here was carried out. JFN would also
like to thank the hospitality of the Max Planck Institut f\"ur
Astrophysik in Garching, where this project was started.  This work
has been supported in part by the National Science Foundation under
grant No. PHY94-07194 to the Institute for Theoretical Physics of the
University of California at Santa Barbara and by the 
Sonderforschungsbereich SFB~375-95 ``Astro--Teilchenphysik'' der 
Deutschen Forchungsgemeinschaft.

\appendix

\section{Shear-corrected artificial viscosity for SPH}
Numerical methods based on a differential formulation of hydrodynamics
conservation laws such as SPH and first-generation finite difference
methods usually rely on artificial viscosity terms in order to deal
with shock-wave discontinuities (see, eg., Potter 1973). Numerically,
these terms aim to mimic the basic physical properties of shocks,
generating entropy at shock fronts and transforming kinetic energy
into internal energy, preferably with minimum post-shock oscillations
in the hydrodynamical variables.  Following von-Neumann \& Richtmyer,
the artificial viscosity is usually restricted to compressive flows
($\nabla\cdot {\bf v} < 0$), and assumed to be proportional to some
power of $\nabla\cdot {\bf v}$. With this definition the artificial
viscosity should vanish in pure shear flows ($\nabla\cdot \vecv = 0$
and $\nabla\times \vecv \ne 0$).

The most widely used formulation of artificial viscosity in SPH is
that of Monaghan and Gingold (1983), who approximate $\nabla \cdot
{\bf v}$ by interparticle velocity differences. The viscous
contribution to the pressure gradient on particle $i$ due to particle
$j$ is given by $m_j Q_{ij} \nabla_i W(|\vecr_i-\vecr_j|,h)$, where
$m_j$ is the mass of particle $j$, $W(r,h)$ is the interpolating
kernel, $h$ is the smoothing length, and $Q_{ij}$ is given by
\begin{eqnarray}
Q_{ij} & = & \left\{
\begin{array}{ll} \frac{-\alpha_v c_{ij}\mu_{ij} +
\beta_v\mu_{ij}^2}{\varrho_{ij}}\, ,
& (\vecr_i - \vecr_j)\cdot (\vecv_i - \vecv_j) \le 0 \\
0\, ,& {\rm otherwise,}
\end{array}
 \right.
  \\
\mu_{ij} & = & \frac{h(\vecv_i-\vecv_j) \cdot
               (\vecr_i-\vecr_j)}{(\vecr_i-\vecr_j)^2 + \eta^2}\, ,
\end{eqnarray}
Here $c_{ij}$ and $\varrho_{ij}$ are the arithmetic means of the sound
velocity $c$ and the density $\varrho$, respectively, at the locations
of particles $i$ and $j$.  The parameter $\eta \approx 0.1h$ is
introduced to prevent numerical divergences. $\alpha_v$ and $\beta_v$
are free parameters that control the amplitude of post-shock
oscillations. For problems involving strong shocks the choice
$\alpha_v=1$ and $\beta_v=2$ is appropriate, but $\alpha_v=0.5$ and
$\beta_v=1$ are also widely used. The Monaghan--Gingold viscosity
gives satisfactory results in test cases, even for strong shocks
(Steinmetz \& M\"uller 1993). Its main disadvantage, however, is that
it does not necessarily vanish in shear-dominated flows, when
$\nabla\cdot\vecv = 0$, but $\nabla\times\vecv\ne 0$. This is because
neighboring particles across the shear surface have a non-zero
velocity component along the line joining their centers. This results
in a non-zero divergence estimate $\mu_{ij}$ and a non-vanishing
viscosity term, $Q_{ij} \ne 0$. This spurious shear viscosity is
especially important when the number of particles, $N$, used is small,
since $Q_{ij} \propto h$ (or $h^2$ depending on which term dominates
in eq.~A1) and typically the smoothing length is chosen so that $h
\propto N^{-1/3}$.

In order to assess the effects of the spurious shear viscosity on the
structure of a centrifugally supported gaseous disk we devised the
following test. A homogeneous, isothermal ($T=10^4$ K) gas sphere is
allowed to collapse in the potential well of a dark matter halo, which
is modeled as a spherically symmetric external potential with a radial
structure given by the density profile proposed by Navarro, Frenk \&
White (1995b) for Cold Dark Matter halos. The total mass of the gas
component is $4 \times 10^{10} M_{\odot}$ and its initial radius is
100 kpc. The dark matter potential is characterized by a circular
speed of $200$ \kms at the virial radius. The gas is initially set in
solid-body rotation with a total angular momentum consistent with a
spin parameter $\lambda=0.15$. This is larger than the typical value
of $0.05$ found in our simulations. Since the gas temperature is well
below the virial temperature of the system, the gas collapses quickly
to form a centrifugally supported disk. Because of the spherical
symmetry assumed for the dark matter potential and the high angular
momentum of the gas (which ensures that the disk is not
self-gravitating) we do not expect non-axisymmetric perturbations to
develop. The disk should remain stationary after the collapse, except
for numerical viscous effects.

A sensitive measure of the effects of shear viscosity is the ratio
between the half-mass radius of the disk and the radius that contains
half its total angular momentum. We denote this ratio by
$R_{JM}$. This is because as shear viscosity transports angular
momentum outwards it also brings material inwards in order to conserve
the total angular momentum. The two upper panels in Figure 11 show the
evolution of $R_{JM}$ when the standard viscosity formulation is used
(eq. A1).  Each run is followed for a total of $1.1 \times 10^{10}$
yrs, slightly less than the age of the universe at $z=0$. Time in this
figure is given in units of the time required to complete a full
rotation at the half-mass radius of the disk ($t_{rot} \sim 6.1 \times
10^8$ yrs). The ratio $R_{JM}$ is normalized to the value just after
the formation of the disk. The different symbols correspond to runs
with different numbers of particles. It is clear from these two panels
that the shear viscosity can alter substantially the structure of the
disk. For $N=247$, $R_{JM}$ halves in about 6 rotations; only for $N$
as large as $4020$ particles the disk remains relatively unchanged
over a Hubble time.

The results presented in the upper panels of Figure 11 are
particularly worrying because, although the disks analyzed in this
paper have $\gsim 4000$ particles at $z=0$, they contain fewer
particles at higher $z$. It is very likely then that the structure of
the disks at $z=0$ can be unduly influenced by the poorer resolution
prevalent at higher redshifts. Motivated by this effect, we decided to
modify the formulation of the artificial viscosity in order to reduce
its shear component. Following Balsara (1995), we define a
``shearfree'' viscosity by
\begin{eqnarray}
\widetilde{Q}_{ij}& = & Q_{ij}\, \frac{f_i + f_j}{2}, \nonumber\\
& & \\
f_i & = & \frac{|\langle\nabla  \cdot \vecv\rangle_i|}{
                |\langle\nabla  \cdot \vecv\rangle_i| +
                |\langle\nabla \times \vecv\rangle_i| +
                0.0001 c_i/h_i}  \, .\nonumber
\end{eqnarray}
The term $0.0001 c_i/h_i$ has been introduced to prevent divergences.
In case of a shear-free, compressive flow ($\nabla\cdot\vecv \ne 0$,
$\nabla\times\vecv = 0$), $f = 1$ and eq.~A3 reverts to the usual
Monaghan \& Gingold formulation (eq.~A1). In the presence of shear
flows, $f<1$ and the importance of the viscous term is reduced. The
suppression factor $f$ is an order of magnitude estimate of the
irrotational component of the flow.  In case of a pure shear flow
($\nabla\cdot\vecv = 0$ and $\nabla\times\vecv \ne 0$), $f$ vanishes
and the viscosity is completely suppressed. (We note that this is not
the only shear-reducing correction that has been
implemented. Hernquist \& Katz (1989) discuss an alternative
formulation. Our tests show, however, that compared with eq.~A3 their
implementation leads to a less sharp definition of shock fronts.)

The effect of implementing such correction on the evolution of disks
is shown in the lower panels of Figure 11. The viscous transport has
been almost completely suppressed, and disks with as few as $247$
particles can be evolved safely for about a Hubble time. Figure 12
summarizes the dependence of viscous effects on $N$ and on the
constants $\alpha_v$ and $\beta_v$. Here we plot the $R_{JM}$
evolutionary timescale ($\tau_{JM}$, measured from Figure 11) as a
function of the number of particles in the disk. Note that $\tau_{JM}$
scales roughly as $N^{-2/3}$, indicating that the $\mu_{ij}^2$ term
dominates in eq.~A1. As expected, $\tau_{JM}$ is inversely
proportional to $\alpha_v$ and $\beta_v$, increasing by about a factor
of 2 when $\alpha_v$ and $\beta_v$ are halved. We conclude that by
using the shearfree formulation of the artificial viscosity (with the
usual values $\alpha_v \approx 1$ and $\beta_v \approx 2$) we can
prevent the viscous angular momentum transport from having an
appreciable effect in our simulations.

\clearpage

\begin{table*}
\begin{center}
\begin{tabular}{crrrrrrrrrr}
Label & $J_{21}^0$ & $\alpha$ & $M_{200}$ & $r_{200}$ & $V_{200}$ &
$N_{gas}$ & $N_{DM}$ & $M_{DM}$ & $M_{gas}$ &$M_{disk}$  \\
      & & & [$10^{10} M_{\odot}$] & [kpc] & [km/s] &
        &  &  [$10^{10} M_{\odot}$] &  [$10^{10} M_{\odot}$]  & [$10^{10} M_{\odot}$]  \\

\tableline
1 & 0.0 &     & 52.0 & 208.5 & 103.7 & 5400 & 5313 & 49.35 & 2.64 & 2.44 \\
$L1$ & 0.0 &     & 56.5 & 214.3 & 106.4 & 1053 & 958 & 53.39 & 3.09 & 2.58 \\
$J1$ & 1.0 & 5.0& 49.93 & 205.6 & 101.6 & 4500 & 5139 & 47.73 & 2.20 & 1.74 \\
&&&&&&&&&& \\
2 & 0.0 &     & 41.2 & 192.8 & 95.8  & 4806 & 4180 & 38.83 & 2.35 & 2.13 \\
$L2$ & 0.0 &     & 41.8 & 194.5 & 96.2  & 805 & 708 & 39.46 & 2.36 & 2.13 \\
$J2$ & 1.0 & 5.0& 38.39 & 188.4 & 93.54 & 4050 & 3920 & 36.41 & 1.98 & 1.65 \\
&&&&&&&&&& \\
3 & 0.0 &     & 25.2 & 162.0 & 81.7  & 2802 & 2561 & 23.79 & 1.37 & 1.31 \\
$L3$ & 0.0 &     & 24.9 & 164.7 & 70.3  & 457 & 422 & 23.52 & 1.34 & 1.28 \\
$J3$ & 1.0 & 5.0& 21.03 & 154.3 & 76.5  & 2311 & 2143 & 19.91 & 1.13 & 0.63 \\
$LJ3$ & 1.0 & 5.0& 24.93 & 165.5 & 80.5  & 368 & 343 & 19.12 & 1.08 & 0.55 \\
&&&&&&&&&& \\
4 & 0.0 &     & 30.6 & 174.6 & 86.7  & 3620 & 3100 & 28.80 & 1.77 & 1.34 \\
$L4$ & 0.0 &     & 34.8 & 182.7 & 90.4  & 678 & 588 & 32.77 & 1.99 & 1.21 \\
$J4$ & 1.0 & 5.0& 30.03 & 173.5 & 86.3  & 3007 & 3075 & 28.56 & 1.47 & 1.05 \\
$LJ4$ & 1.0 & 5.0& 33.37 & 180.0 & 89.4  & 549 & 570 & 31.77 & 1.61 & 1.07 \\
$J4a$ &10.0& 1.0& 30.20 & 173.9 & 86.5  & 2659 & 3112 & 28.91 & 1.30 & 0.744 \\
&&&&&&&&&& \\
5 & 0.0 &     & 234.3 & 344.0 & 171.1& 4410 & 5217 & 224.3 & 9.98 & 8.82 \\
$L5$ & 0.0 &     & 228.9 & 347.9 & 168.2& 765 & 847 & 218.6 & 10.39 & 9.05 \\
$J5$ & 1.0 & 5.0& 230.7 & 342.3 & 170.3 & 3927 & 5159 & 221.9 & 8.89 & 4.89 \\
$LJ5$ & 1.0 & 5.0& 227.5 & 341.1 & 169.4 & 617 & 849 & 219.1 & 8.38 & 6.34 \\
$J5a$ &10.0& 1.0& 233.1 & 343.5 & 170.9 & 4900 & 5163 & 222.0 & 11.1 & 5.03 \\
&&&&&&&&&& \\
6 & 0.0 &     & 212.7 & 333.4 & 165.7& 4085 & 4731 & 203.5 & 9.25 & 7.59 \\
$L6$ & 0.0 &     & 229.8 & 342.6 & 169.9& 843 & 846 & 218.3 & 11.45 & 9.15 \\
$J6$ & 1.0 & 5.0& 213.7 & 336.0 & 165.4 & 3654 & 4776 & 205.4 & 8.27 & 5.45 \\
$LJ6$ & 1.0 & 5.0& 225.4 & 342.6 & 168.2 & 695 & 837 & 216.0 & 9.44 & 6.59 \\
&&&&&&&&&& \\
7 & 0.0 &     & 336.3 & 388.2 & 192.9& 7255 & 7439 & 319.9 &16.42 & 13.41 \\
$L7$ & 0.0 &     & 383.4 & 405.9 & 201.6& 1328 & 1416& 365.4 & 18.04 & 14.06 \\
$J7$ & 1.0 & 5.0& 342.4 & 393.1 & 193.5 & 6495 & 7620 & 327.7 & 14.7 & 9.31 \\
$LJ7$ & 1.0 & 5.0& 381.8 & 407.7 & 200.8 & 1117 & 1421& 366.7 & 15.17& 6.76 \\
&&&&&&&&&& \\
8 & 0.0 &     & 253.3 & 353.1 & 175.7& 5669 & 5593 & 240.5 &12.83 & 9.80 \\
$L8$ & 0.0 &     & 208.8 & 334.2 & 163.8& 798 & 767 & 197.9 & 10.84 & 9.71 \\
$J8$ & 1.0 & 5.0& 240.3 & 349.3 & 171.9 & 5084 & 5530 & 229.2 & 11.1 & 5.61 \\
$LJ8$ & 1.0 & 5.0& 210.4 & 334.3 & 164.5 & 613 & 783 & 202.0 & 8.32 & 6.45 \\
&&&&&&&&&& \\

\end{tabular}
\end{center}

\caption{Parameters of the numerical experiments.} \label{tbl-1}

\end{table*}

\clearpage

\begin{figure}
\plotone{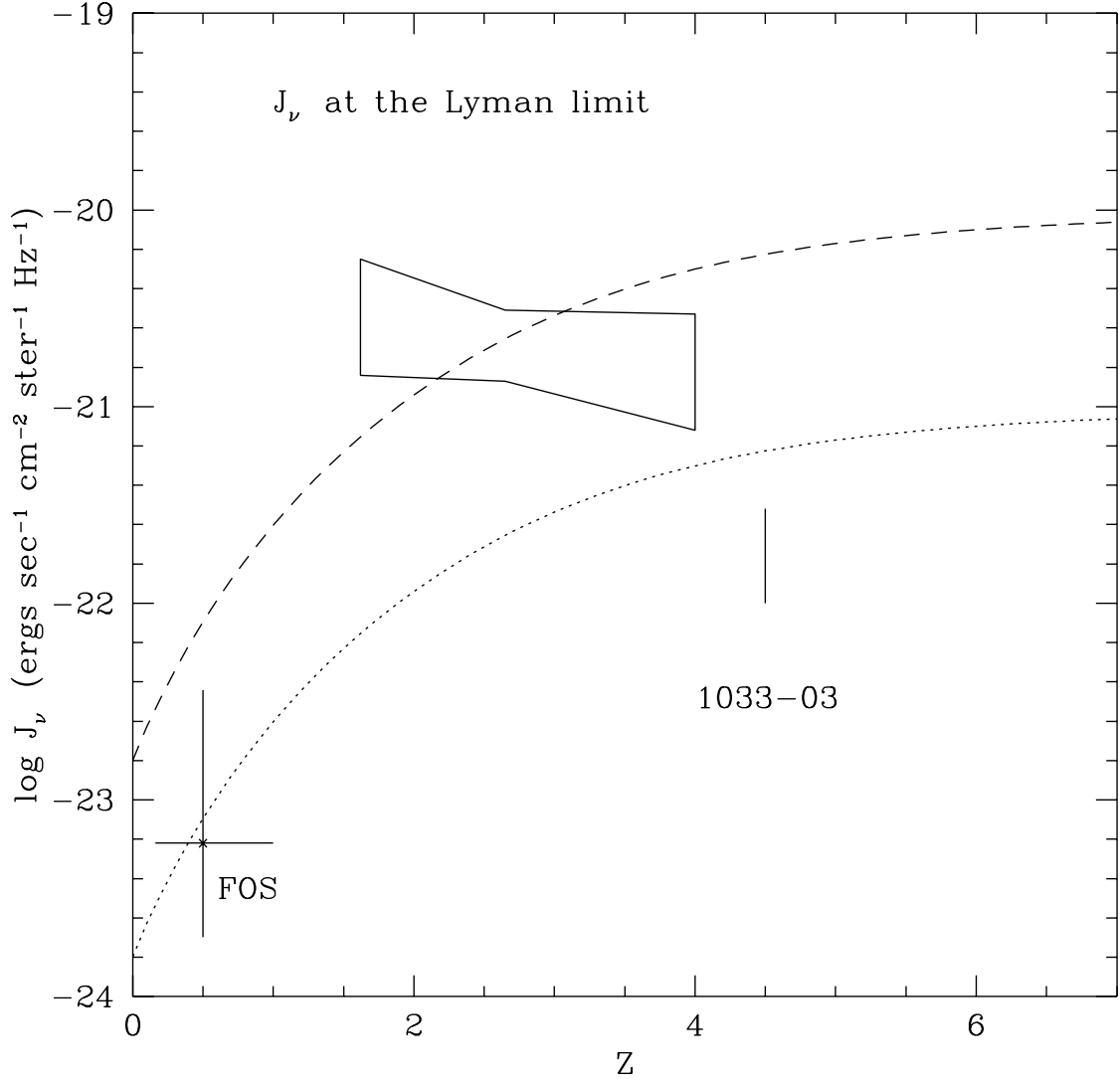}
\caption{
The redshift evolution of the photoionizing flux parameter
$J_{21}(z)$. The lower curve corresponds to our fiducial choice of UV
background (runs $J$). The upper curve represents an extreme version
of the UV field (runs $Ja$). The symbols are observational constraints
taken from the review by Bechtold (1995).}
\end{figure}

\begin{figure}
\plotone{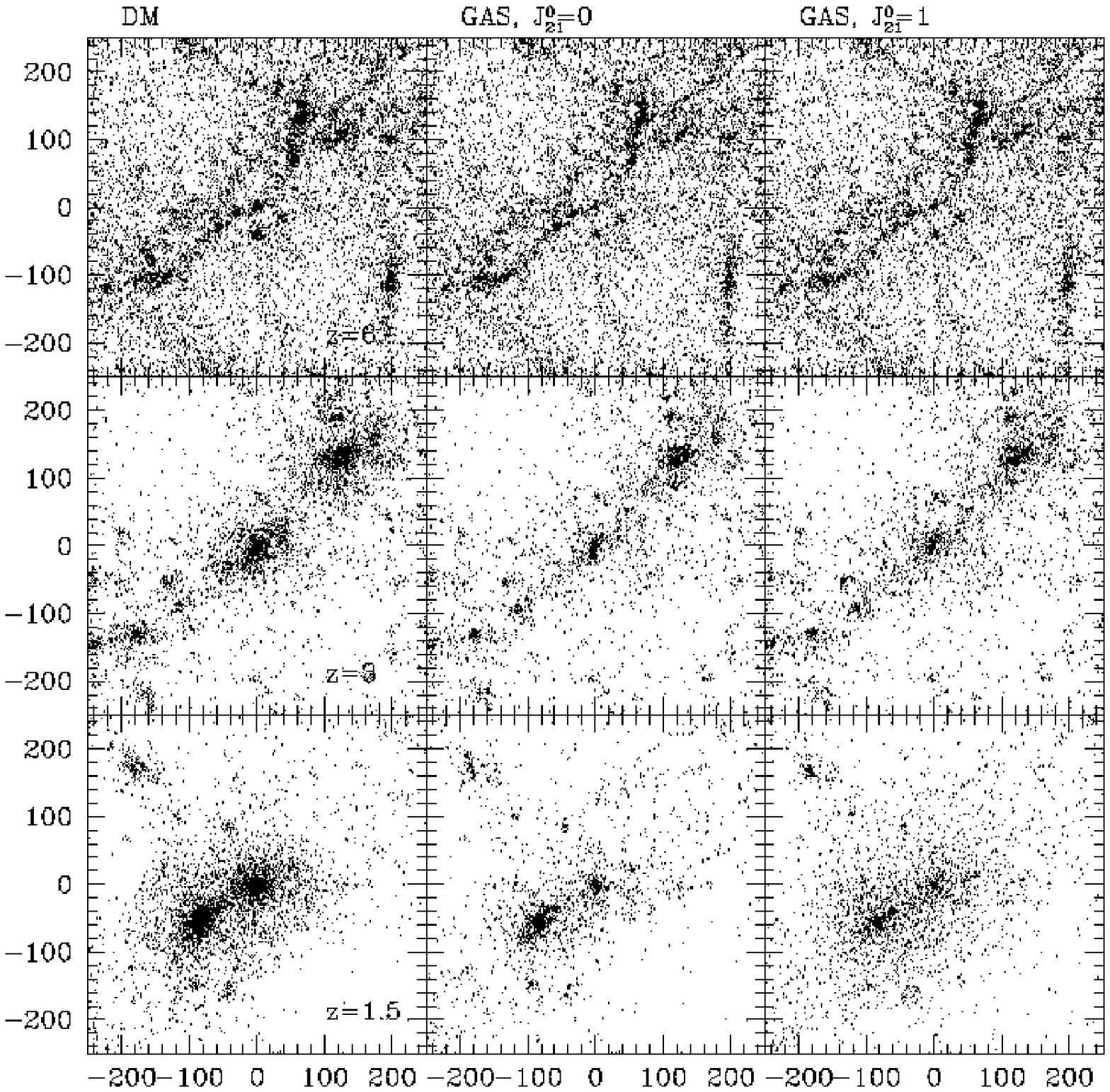}
\figurenum{2}
\caption{}
\end{figure}

\begin{figure}
\plotone{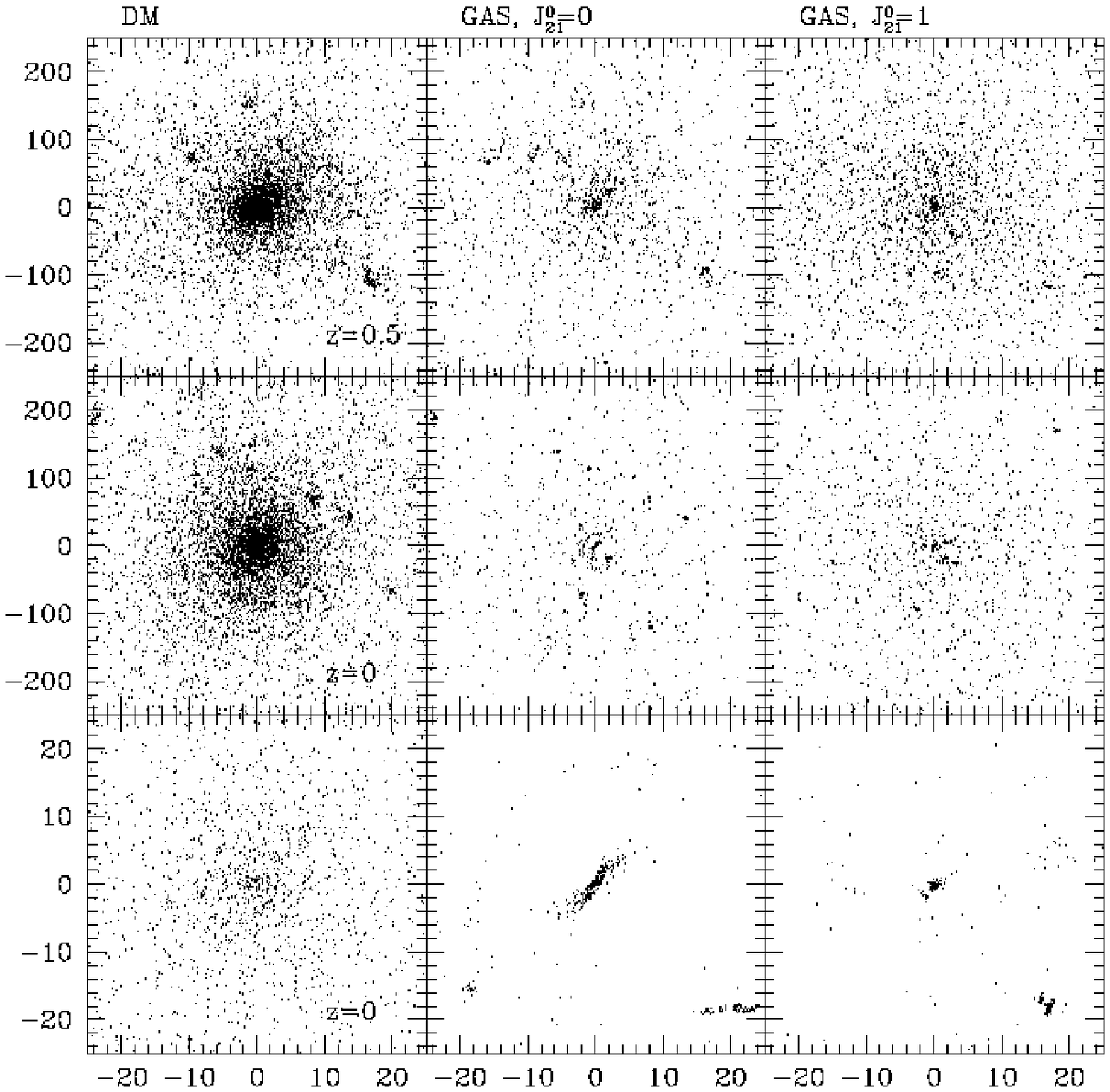}
\figurenum{2-cont}
\caption{
Dark matter and gas particles corresponding to run $5$, plotted at
different times. Boxes are $500$ physical kpc on a side, except for
the last row, which shows a blow-up of the particle distribution at
$z=0$. The box in this case is $50$ kpc on a side. The left column
corresponds to the dark matter, the middle column to gas in the
absence of photoionization effects, and the right column to gas
including the effects of our fiducial UV background.}
\end{figure}

\begin{figure}
\plotone{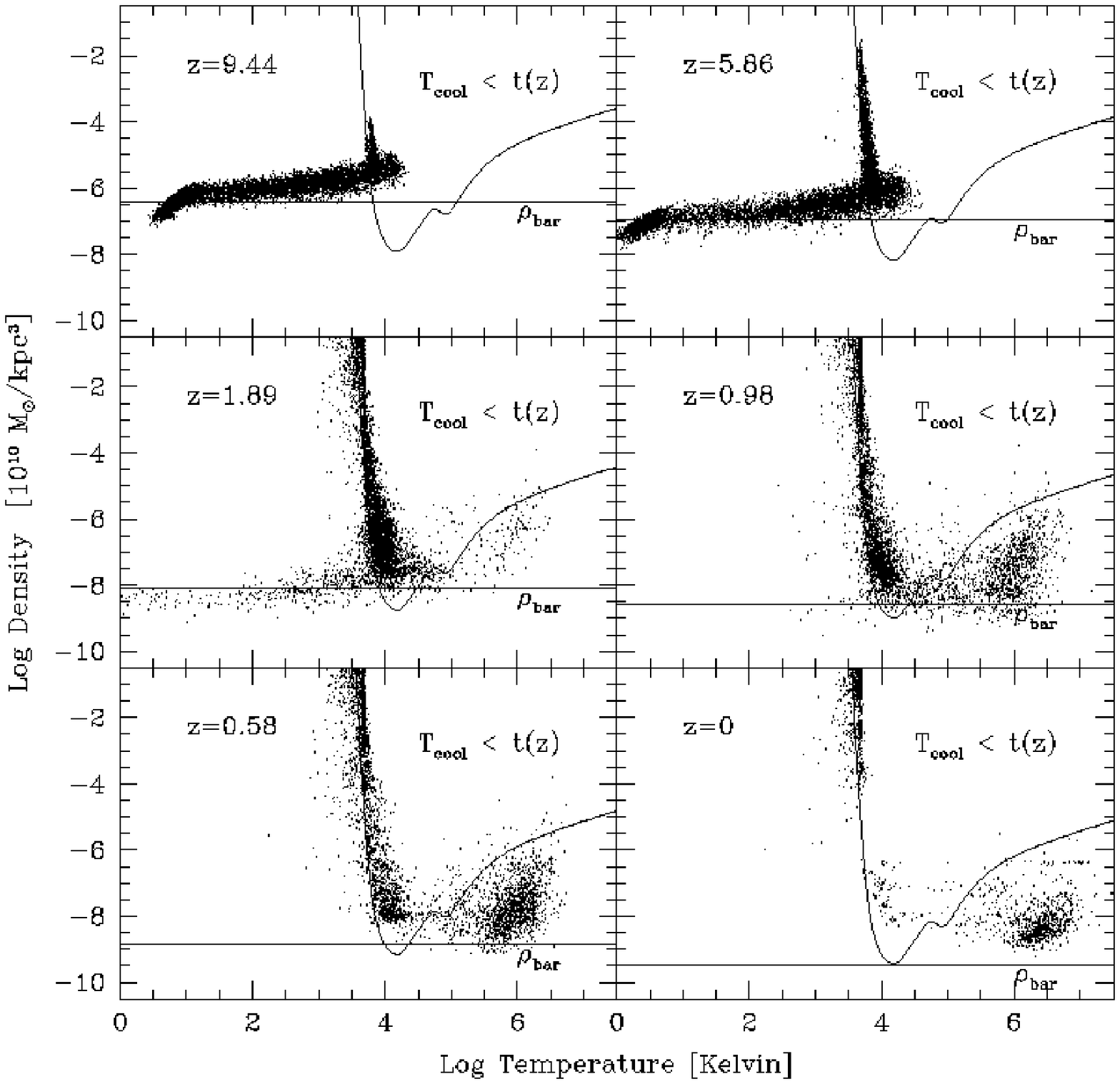}
\figurenum{3}
\caption{
Distribution in the density-temperature plane of the gas particles
found within the virial radius of the system at $z=0$ in a run without
photoionization. Corresponding redshifts are labeled in each
panel. The horizontal line represents the mean baryon density. The
solid curve separates the region where the cooling times are shorter
and longer than the age of the universe. Shock-heated particles move
along nearly horizontal lines in this plot. Note that only a small
fraction of the particles get shock heated to $T \sim 10^6$ K, the
virial temperature of the halo. The rest cool very quickly as soon as
they reach $10^4$ K and collapse to form rotationally supported
disks.}
\end{figure}

\begin{figure}
\plotone{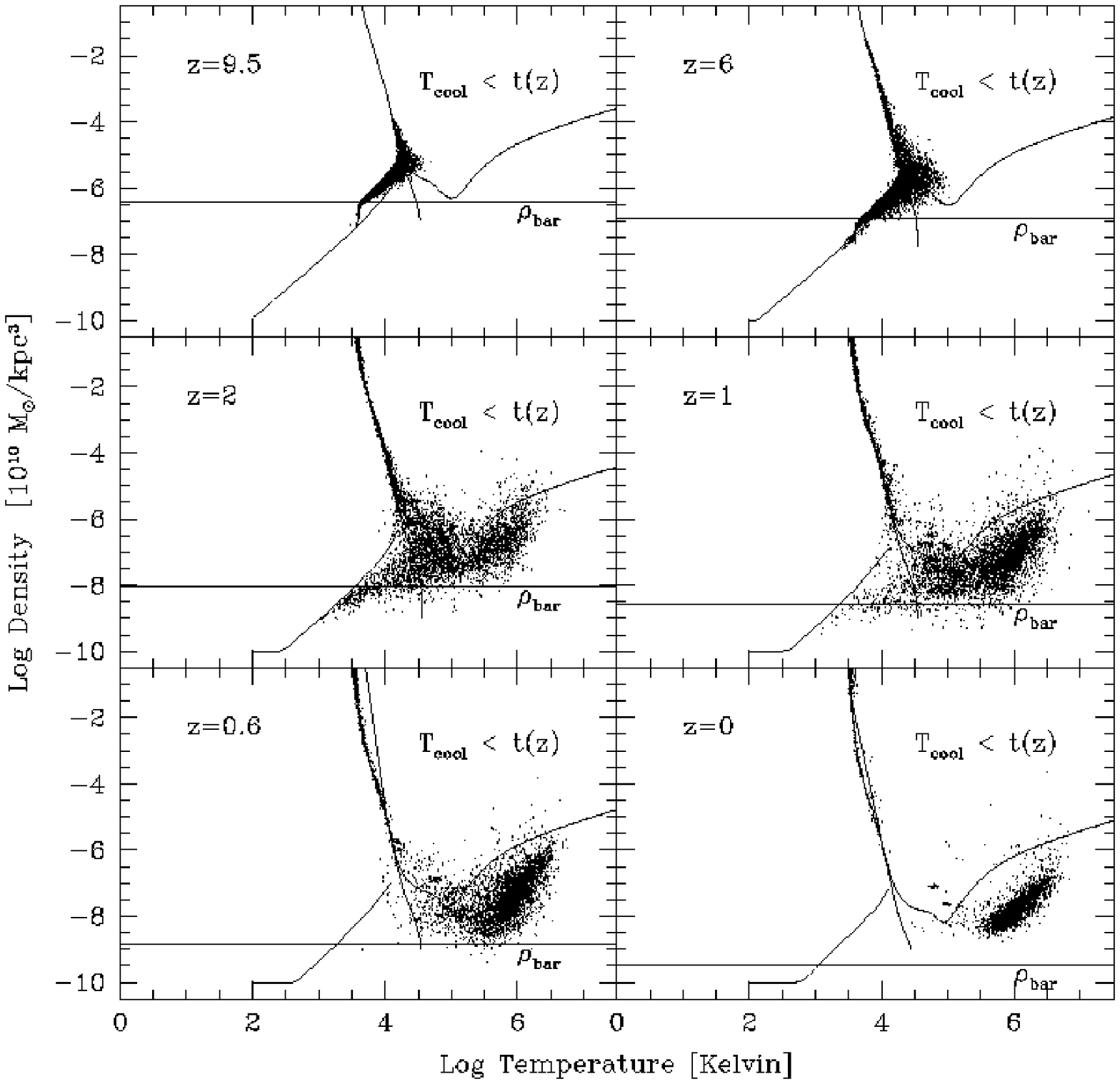}
\figurenum{4}
\caption{
Same as Figure 3, but for our fiducial UV background choice. The
horizontal curve and the curve on the upper right of each panel are as
in Figure 3. The almost diagonal curve at the lower left indicates
where the photoheating timescale equals the age of the universe. The
almost vertical curve indicates the equilibrium temperature; ie. the
loci in this plane where cooling and heating rates are the same. Note
that at high redshift the gas has been photoheated to about $10^4$
K. Cooling rates have also been dramatically reduced, especially at $T
< 10^5$ K. Because of the lowered cooling rates, a larger fraction of
particles can be shock heated to the virial temperature of the halo,
$T_{vir} \sim 10^6$ K. Most of the particles that reach that
temperature remain until $z=0$ in a hot, pressure supported, diffuse
halo because of the long cooling times characteristic of that
temperature.}
\end{figure}

\begin{figure}
\plotone{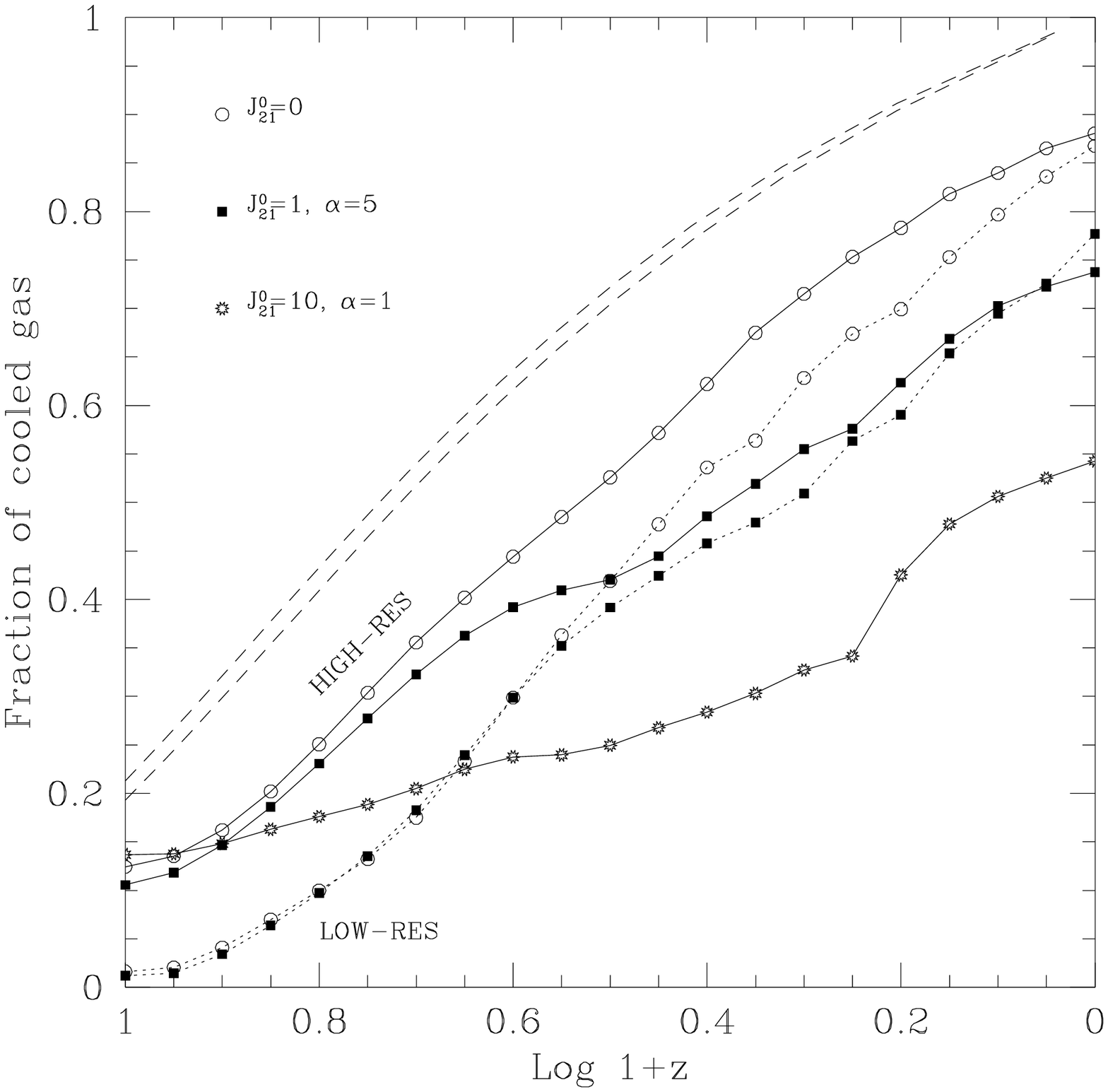}
\figurenum{5}
\caption{
The evolution of the gas mass in dense, cold clumps ($\rho > 10^5
M_{\odot}/$kpc$^3$), expressed as a fraction of the final gas mass of
the system. The solid lines show the average of all
``high-resolution'' runs with the same UV parameters. Open circles are
used for runs without UV background, solid squares for our fiducial
choice of UV radiation field ($J_{21}^0=1$, $\alpha=5$), and starred
symbols correspond to an extreme version of the photoionizing
background ($J_{21}^0=10$, $\alpha=1$). The dotted lines correspond to
the ``low-resolution'' runs (runs $L$ and $LJ$). The dashed lines show
the fraction of mass in clumps with virial temperatures larger than
$10^4$ K, computed using the Press-Schechter theory. This represents a
theoretical upper bound to the fraction of mass that can cool at each
redshift.}
\end{figure}

\begin{figure}
\plotone{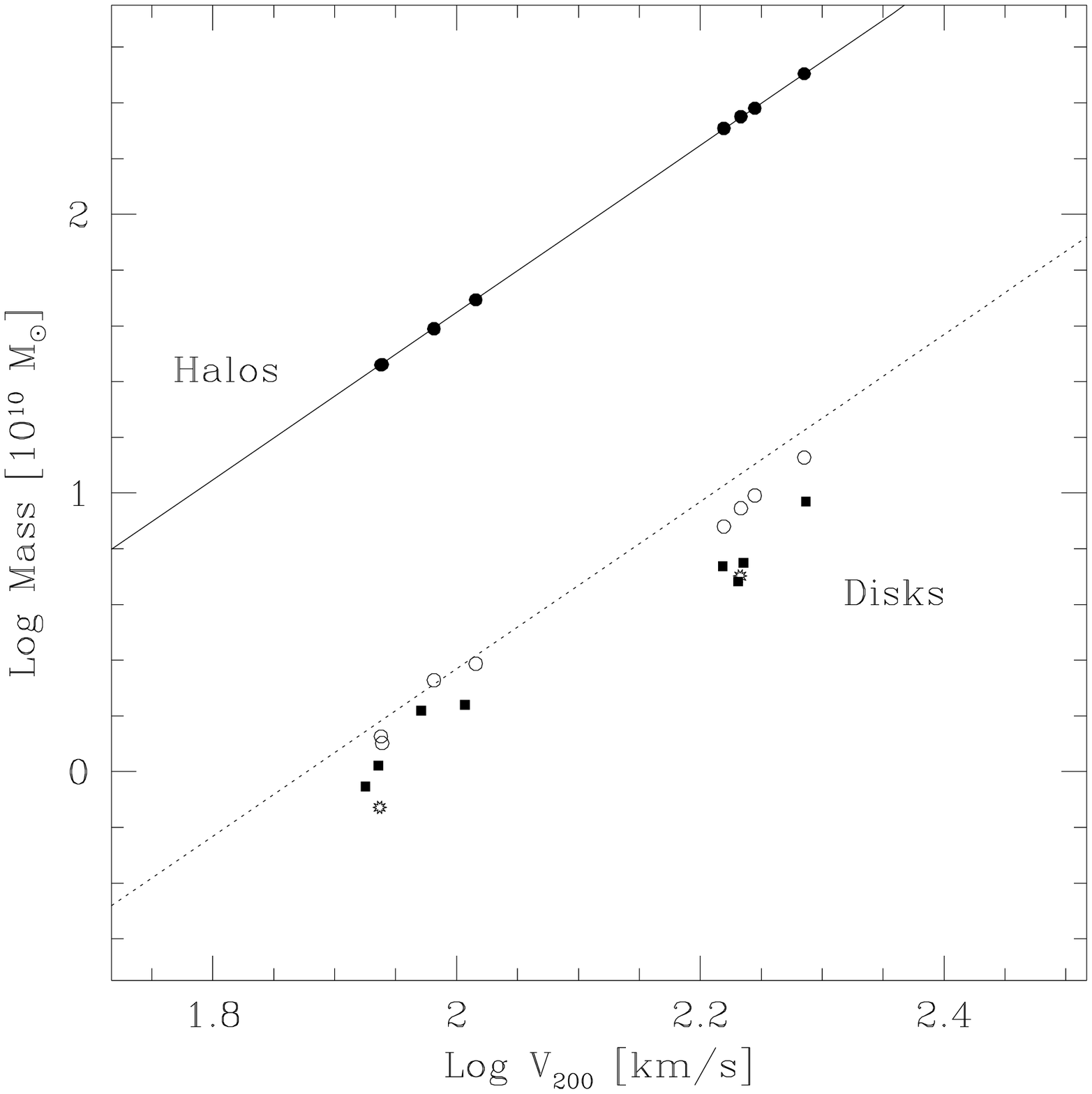}
\figurenum{6}
\caption{
The mass of the system in the dark halo and in the central gaseous
disk, as a function of the circular velocity at the virial radius. The
halo mass agrees very well with the relation $M_{DM} = (1-\Omega_b)
M_{200} \propto V_{200}^3$ (solid line). The dotted line indicates
where the central gaseous disks would fall if all the gaseous mass of
the system had collapsed to the center ($M_{disk}= \Omega_b M_{200}
\propto V_{200}^3$). Open circles correspond to runs without UV
background; solid squares correspond to our fiducial background choice
($J_{21}^0=1$, $\alpha=5$). Starred symbols correspond to runs with an
extermely energetic background ($J_{21}^0=10$, $\alpha=1$).}
\end{figure}

\begin{figure}
\plotone{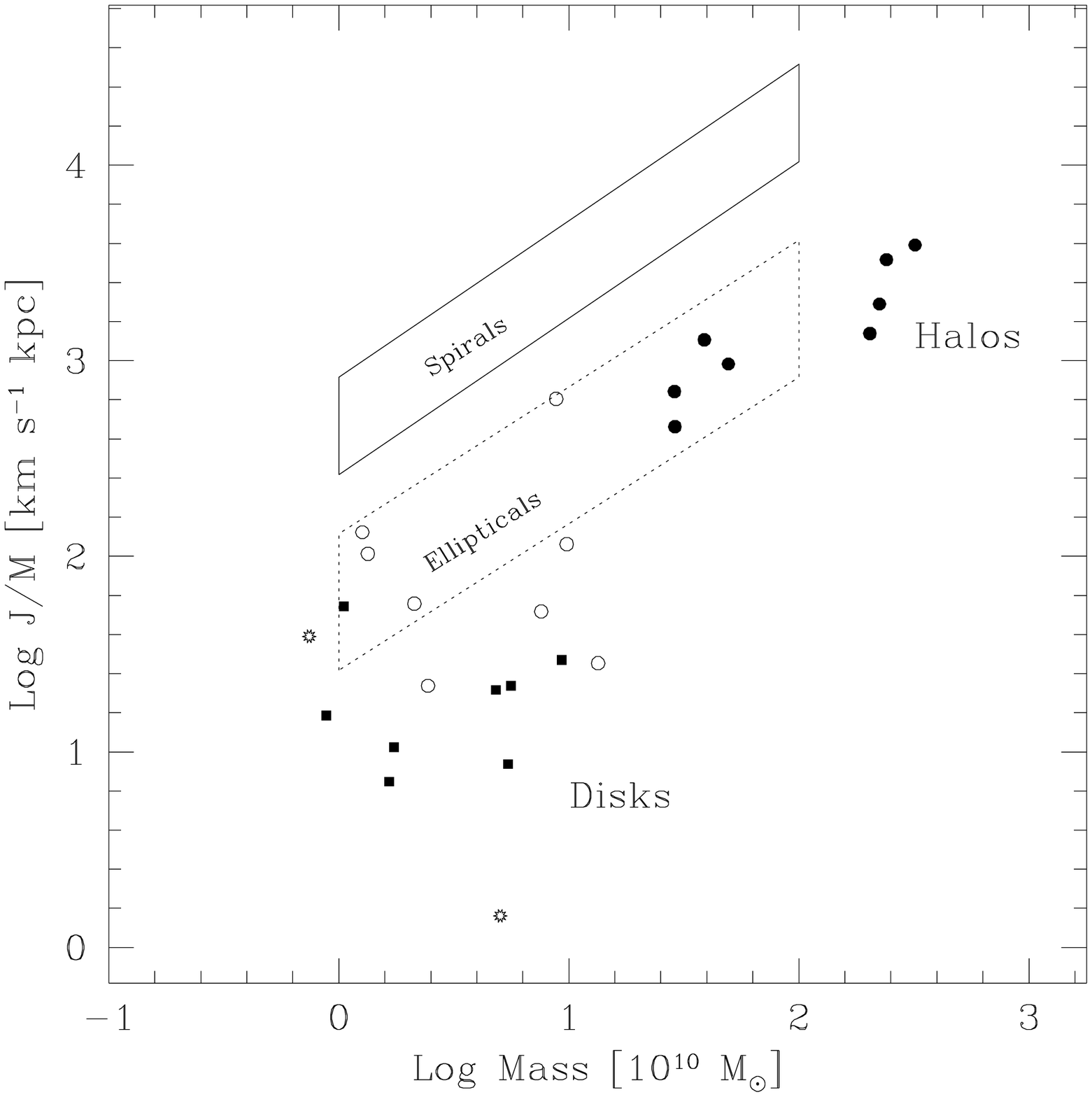}
\figurenum{7}
\caption{
The specific angular momentum of dark halos and gaseous disks, as a
function of mass. The boxes enclose the region occupied by spiral and
elliptical galaxies, as given by Fall (1983). Symbols are as in Figure
6. Note that the halos' $J/M$ scale approximately as $M^{2/3}$, as
expected if all systems had the same value of the rotation parameter
$\lambda$ (see text for a definition). Gaseous disks have much lower
angular momenta than observed spirals, a consequence of the role of
mergers during the assembly of the disks. Note that the inclusion of a
UV radiation field seems to aggravate this problem.}
\end{figure}

\begin{figure}
\plotone{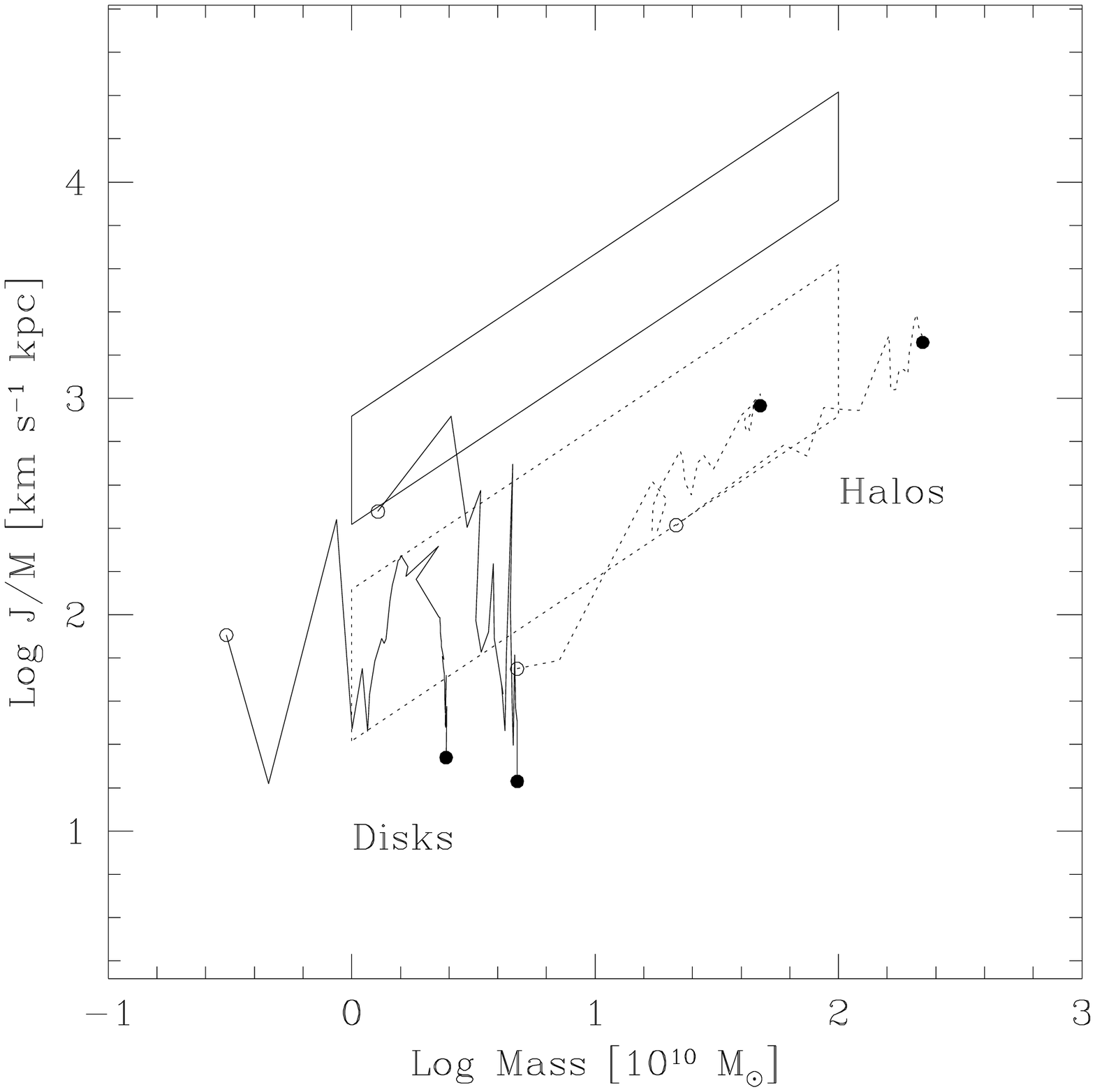}
\figurenum{8}
\caption{
Evolution of the dark halo and central gaseous disk in the $J/M$ vs
$M$ plane, from $z=5$ (open circles) to $z=0$ (solid circles). Boxes
are as in Figure 7. The evolution of two systems is shown. Note that
at $z=5$ the gas and dark matter appear to have the same specific
angular momentum. This is a direct consequence of numerical
limitations, which make the poorly resolved collapse of high-redshift
halos artificially smooth. The mass of the system grows steadily by
mergers, which are accompanied by an increase in the spin of the halo
and a decrease in the spin of the central disk. The latter results
from angular momentum being transferred from the gas to the halo
during mergers.}
\end{figure}

\begin{figure}
\plotone{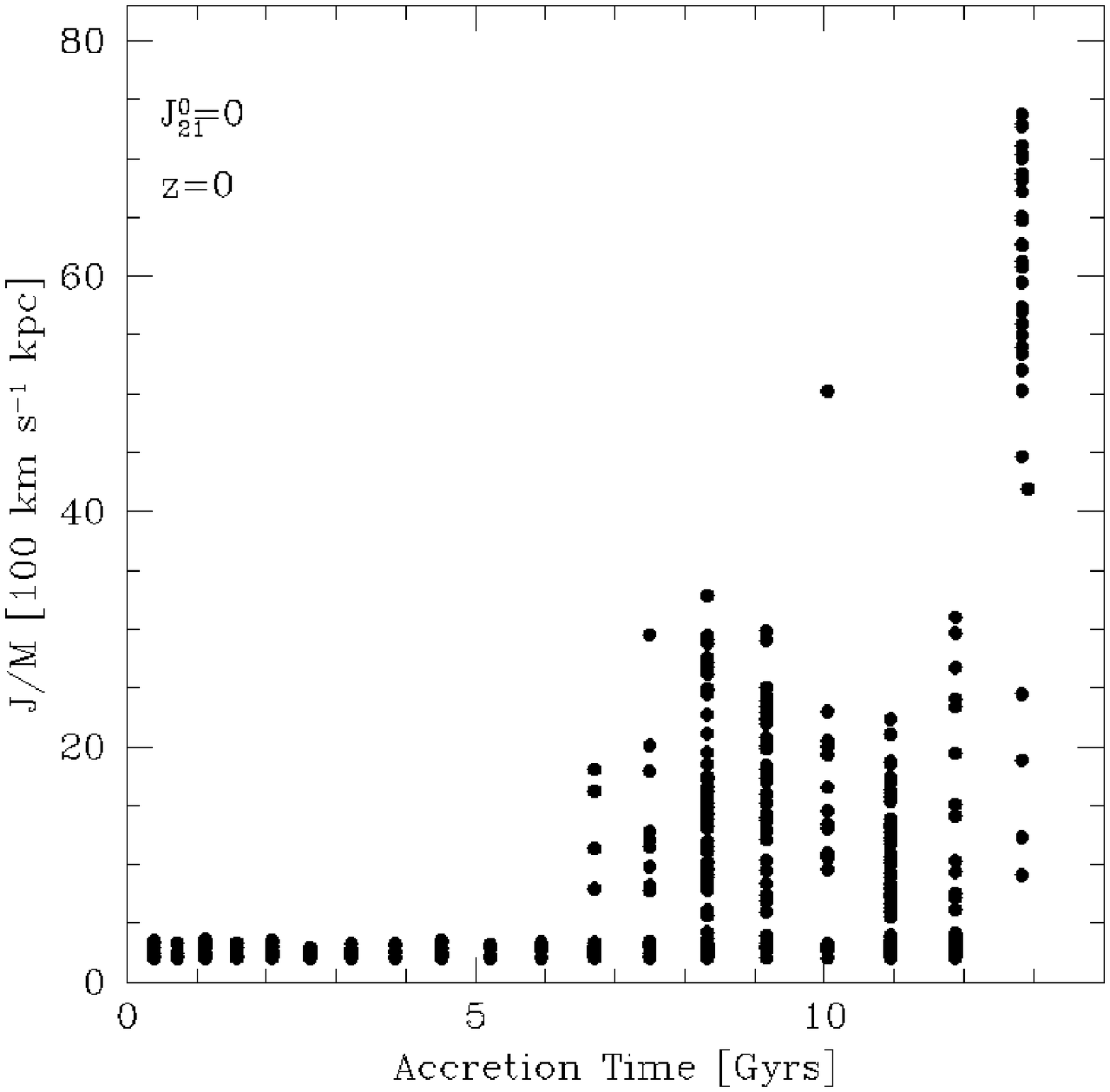}
\figurenum{9}
\caption{
The angular momentum of gas particles in the disk at $z=0$ plotted as
a function of the time at which it was accreted into the disk (run
$5$, no UV background). Note that particles accreted late have higher
specific angular momentum, indicating that the outer regions of the
disk are assembled later (ie. the disk grows
``inside-out''). Suppression of late accretion would result in a
reduction of the disk spin. This is the reason why the presence of UV
radiation seems to lower the spin of gaseous disks. }
\end{figure}

\begin{figure}
\plotone{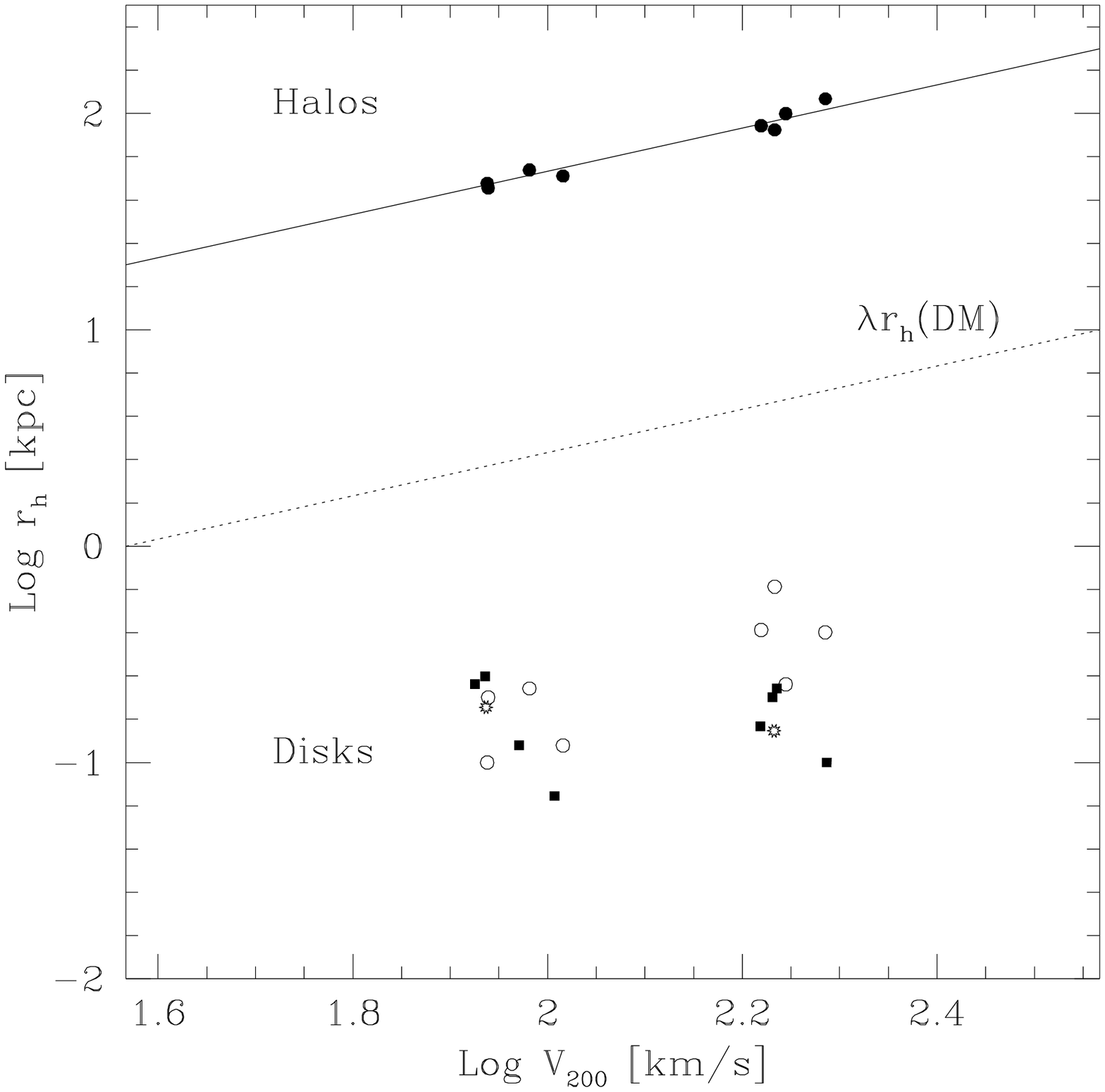}
\figurenum{10}
\caption{
The half-mass radius of dark matter halos and gaseous disks as a
function of the circular speed of the system. The solid line is a fit
to the halo data of the form $r_h \propto r_{200} \propto
V_{200}$. The good agreement between this fit and the data indicates
that the spatial structure of the dark matter halos is similar and
roughly independent of mass.  The half mass radii of the disks are
much smaller than expected if both halos and disks had the same
specific angular momentum. This is indicated by the dotted line, which
is parallel to the solid line but displaced vertically by a factor
$\lambda=0.05$, the typical value of the rotation parameter. Note that
the dotted line is a good approximation to the sizes of spiral disks.}
\end{figure}

\begin{figure}
\plotone{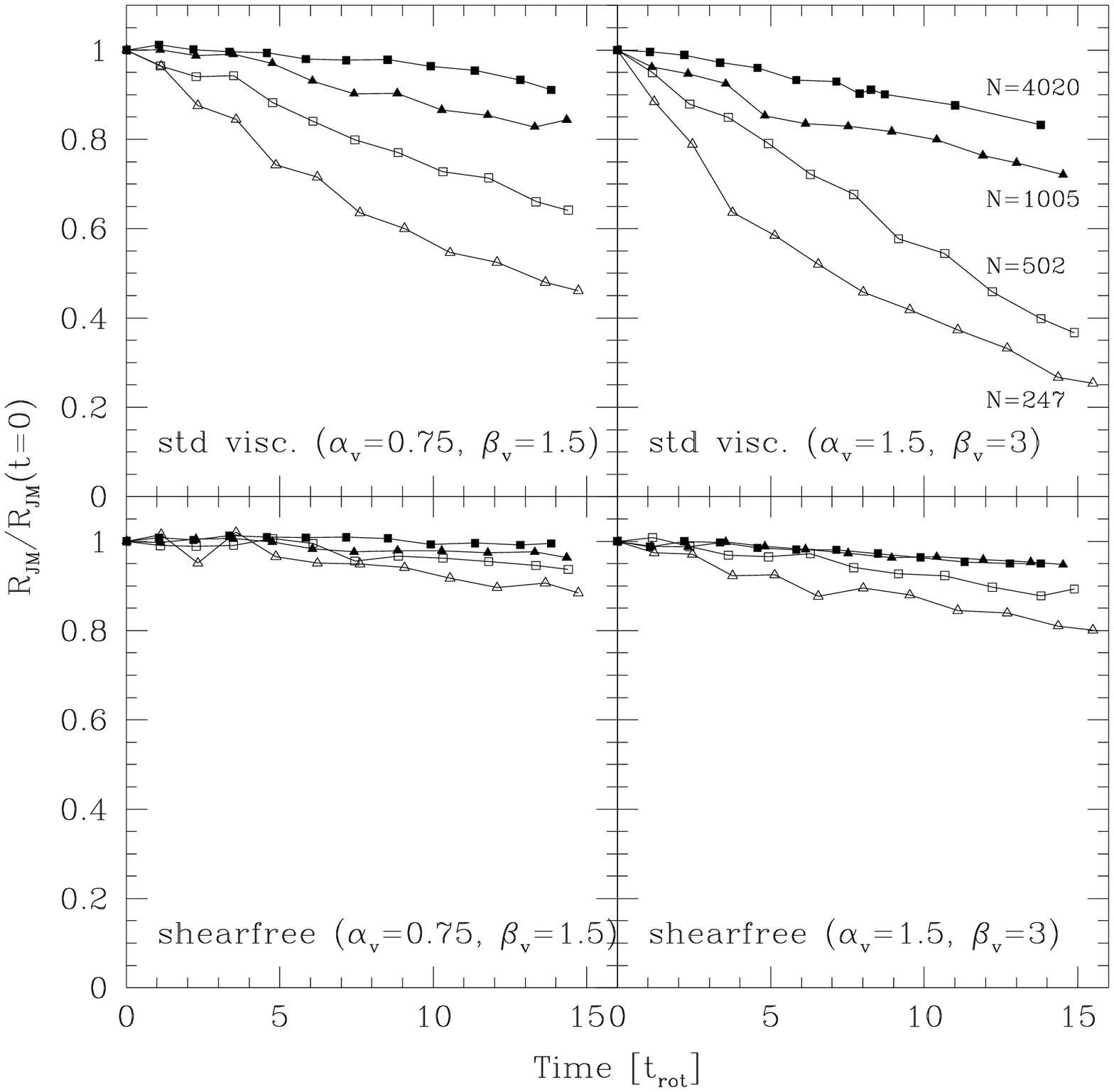}
\figurenum{11}
\caption{
Time evolution of the ratio between half-mass and half-angular
momentum radii. Time is given in units of a full rotation of the disk
($t_{rot} \sim 6 \times 10^8 $ yrs). $R_{JM}$ is normalized to its
value just after the disk is assembled.}
\end{figure}

\begin{figure}
\plotone{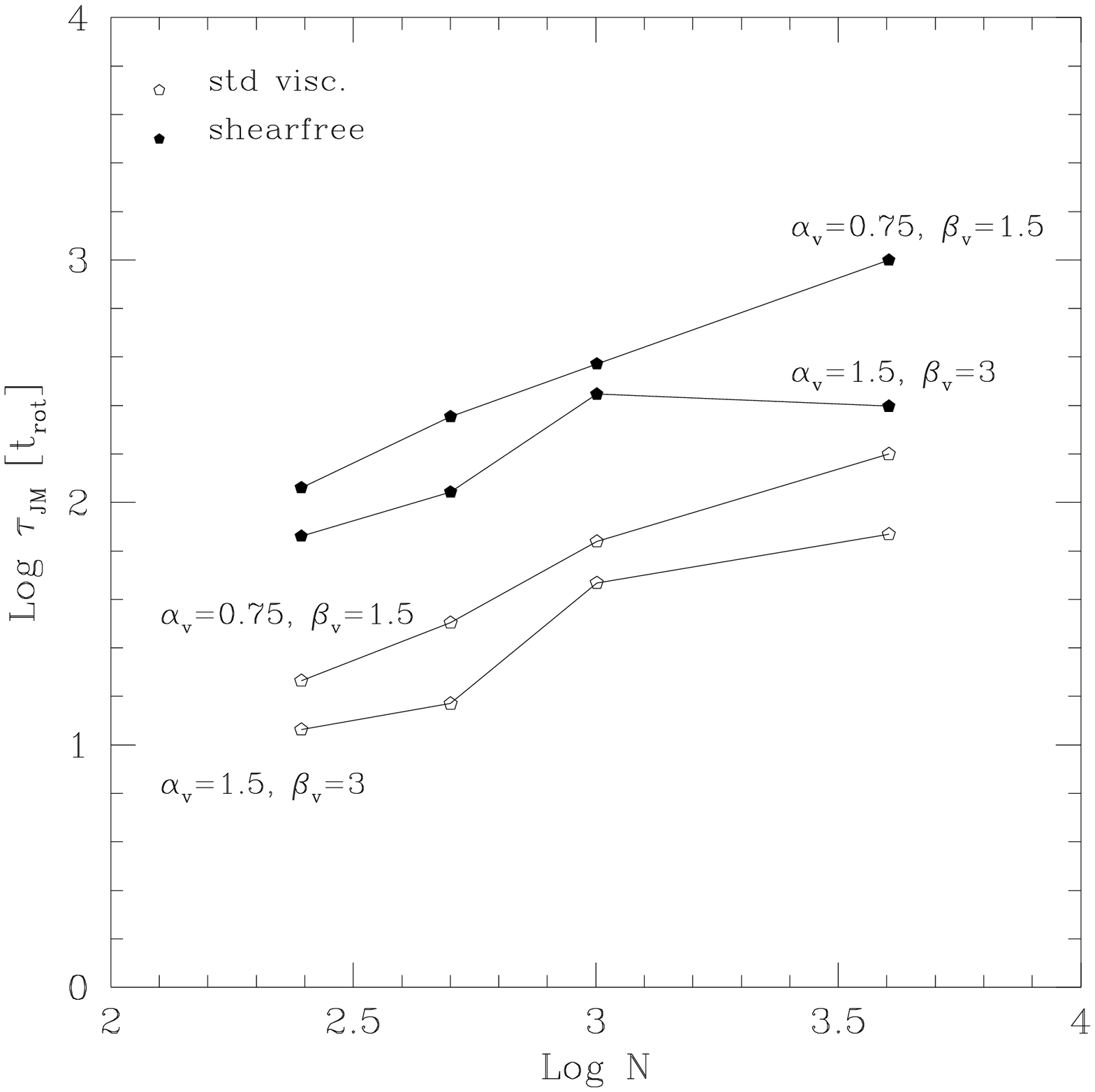}
\figurenum{12}
\caption{
Timescale for artificial angular momentum transport as a function of
particle number $N$ for different values of the viscosity constants.}
\end{figure}

\end{document}